\documentclass[12pt,prd,showpacs,tightenlines,nofootinbib]{revtex4}
\usepackage{bm}
\usepackage{graphics}
\usepackage{rotating}
\usepackage{epsfig}
\begin{document}
\begin{flushright}
{\bf Preprint SSU-HEP-04/08\\
Samara State University\\
version 1}
\end{flushright}
\title{Nuclear structure corrections
to gyromagnetic factor of the bound lepton
\footnote{Talk presented on poster session of the Bogolubov Conference
"Problems of Theoretical and Mathematical Physics", Moscow-Dubna, September
2-6, 2004}}
\author {A.P.Martynenko}
\affiliation{Samara State University,
443011, Samara, Pavlov 1, Russia}
\begin{abstract}
In the framework of the quasipotential method the covariant expression for the
two-particle vertex operator is obtained. The nuclear structure
corrections of orders $(Z\alpha)^4$, $(Z\alpha)^5$ including
recoil effects to gyromagnetic factors of the bound electron and muon
are calculated. Numerical value of the contribution of order
$(Z\alpha)^5$ is obtained by means of the dipole parameterization
for the nuclear charge form factor in the range of the nuclear charges
$Z=7\div 32$.
\end{abstract}


\maketitle

\immediate\write16{<<WARNING: LINEDRAW macros work with emTeX-dvivers
                    and other drivers supporting emTeX \special's
                    (dviscr, dvihplj, dvidot, dvips, dviwin, etc.) >>}

\newdimen\Lengthunit       \Lengthunit  = 1.5cm
\newcount\Nhalfperiods     \Nhalfperiods= 9
\newcount\magnitude        \magnitude = 1000

\catcode`\*=11
\newdimen\L*   \newdimen\d*   \newdimen\d**
\newdimen\dm*  \newdimen\dd*  \newdimen\dt*
\newdimen\a*   \newdimen\b*   \newdimen\c*
\newdimen\a**  \newdimen\b**
\newdimen\xL*  \newdimen\yL*
\newdimen\rx*  \newdimen\ry*
\newdimen\tmp* \newdimen\linwid*

\newcount\k*   \newcount\l*   \newcount\m*
\newcount\k**  \newcount\l**  \newcount\m**
\newcount\n*   \newcount\dn*  \newcount\r*
\newcount\N*   \newcount\*one \newcount\*two  \*one=1 \*two=2
\newcount\*ths \*ths=1000
\newcount\angle*  \newcount\q*  \newcount\q**
\newcount\angle** \angle**=0
\newcount\sc*     \sc*=0

\newtoks\cos*  \cos*={1}
\newtoks\sin*  \sin*={0}

\catcode`\[=13

\def\rotate(#1){\advance\angle**#1\angle*=\angle**
\q**=\angle*\ifnum\q**<0\q**=-\q**\fi
\ifnum\q**>360\q*=\angle*\divide\q*360\multiply\q*360\advance\angle*-\q*\fi
\ifnum\angle*<0\advance\angle*360\fi\q**=\angle*\divide\q**90\q**=\q**
\def\sgcos*{+}\def\sgsin*{+}\relax
\ifcase\q**\or
 \def\sgcos*{-}\def\sgsin*{+}\or
 \def\sgcos*{-}\def\sgsin*{-}\or
 \def\sgcos*{+}\def\sgsin*{-}\else\fi
\q*=\q**
\multiply\q*90\advance\angle*-\q*
\ifnum\angle*>45\sc*=1\angle*=-\angle*\advance\angle*90\else\sc*=0\fi
\def[##1,##2]{\ifnum\sc*=0\relax
\edef\cs*{\sgcos*.##1}\edef\sn*{\sgsin*.##2}\ifcase\q**\or
 \edef\cs*{\sgcos*.##2}\edef\sn*{\sgsin*.##1}\or
 \edef\cs*{\sgcos*.##1}\edef\sn*{\sgsin*.##2}\or
 \edef\cs*{\sgcos*.##2}\edef\sn*{\sgsin*.##1}\else\fi\else
\edef\cs*{\sgcos*.##2}\edef\sn*{\sgsin*.##1}\ifcase\q**\or
 \edef\cs*{\sgcos*.##1}\edef\sn*{\sgsin*.##2}\or
 \edef\cs*{\sgcos*.##2}\edef\sn*{\sgsin*.##1}\or
 \edef\cs*{\sgcos*.##1}\edef\sn*{\sgsin*.##2}\else\fi\fi
\cos*={\cs*}\sin*={\sn*}\global\edef\gcos*{\cs*}\global\edef\gsin*{\sn*}}\relax
\ifcase\angle*[9999,0]\or
[999,017]\or[999,034]\or[998,052]\or[997,069]\or[996,087]\or
[994,104]\or[992,121]\or[990,139]\or[987,156]\or[984,173]\or
[981,190]\or[978,207]\or[974,224]\or[970,241]\or[965,258]\or
[961,275]\or[956,292]\or[951,309]\or[945,325]\or[939,342]\or
[933,358]\or[927,374]\or[920,390]\or[913,406]\or[906,422]\or
[898,438]\or[891,453]\or[882,469]\or[874,484]\or[866,499]\or
[857,515]\or[848,529]\or[838,544]\or[829,559]\or[819,573]\or
[809,587]\or[798,601]\or[788,615]\or[777,629]\or[766,642]\or
[754,656]\or[743,669]\or[731,681]\or[719,694]\or[707,707]\or
\else[9999,0]\fi}

\catcode`\[=12

\def\GRAPH(hsize=#1)#2{\hbox to #1\Lengthunit{#2\hss}}

\def\Linewidth#1{\global\linwid*=#1\relax
\global\divide\linwid*10\global\multiply\linwid*\mag
\global\divide\linwid*100\special{em:linewidth \the\linwid*}}

\Linewidth{.4pt}
\def\sm*{\special{em:moveto}}
\def\sl*{\special{em:lineto}}
\let\moveto=\sm*
\let\lineto=\sl*
\newbox\spm*   \newbox\spl*
\setbox\spm*\hbox{\sm*}
\setbox\spl*\hbox{\sl*}

\def\mov#1(#2,#3)#4{\rlap{\L*=#1\Lengthunit
\xL*=#2\L* \yL*=#3\L*
\xL*=\xscale\xL* \yL*=\yscale\yL*
\rx* \the\cos*\xL* \tmp* \the\sin*\yL* \advance\rx*-\tmp*
\ry* \the\cos*\yL* \tmp* \the\sin*\xL* \advance\ry*\tmp*
\kern\rx*\raise\ry*\hbox{#4}}}

\def\rmov*(#1,#2)#3{\rlap{\xL*=#1\yL*=#2\relax
\rx* \the\cos*\xL* \tmp* \the\sin*\yL* \advance\rx*-\tmp*
\ry* \the\cos*\yL* \tmp* \the\sin*\xL* \advance\ry*\tmp*
\kern\rx*\raise\ry*\hbox{#3}}}

\def\lin#1(#2,#3){\rlap{\sm*\mov#1(#2,#3){\sl*}}}

\def\arr*(#1,#2,#3){\rmov*(#1\dd*,#1\dt*){\sm*
\rmov*(#2\dd*,#2\dt*){\rmov*(#3\dt*,-#3\dd*){\sl*}}\sm*
\rmov*(#2\dd*,#2\dt*){\rmov*(-#3\dt*,#3\dd*){\sl*}}}}

\def\arrow#1(#2,#3){\rlap{\lin#1(#2,#3)\mov#1(#2,#3){\relax
\d**=-.012\Lengthunit\dd*=#2\d**\dt*=#3\d**
\arr*(1,10,4)\arr*(3,8,4)\arr*(4.8,4.2,3)}}}

\def\arrlin#1(#2,#3){\rlap{\L*=#1\Lengthunit\L*=.5\L*
\lin#1(#2,#3)\rmov*(#2\L*,#3\L*){\arrow.1(#2,#3)}}}

\def\dasharrow#1(#2,#3){\rlap{{\Lengthunit=0.9\Lengthunit
\dashlin#1(#2,#3)\mov#1(#2,#3){\sm*}}\mov#1(#2,#3){\sl*
\d**=-.012\Lengthunit\dd*=#2\d**\dt*=#3\d**
\arr*(1,10,4)\arr*(3,8,4)\arr*(4.8,4.2,3)}}}

\def\clap#1{\hbox to 0pt{\hss #1\hss}}

\def\ind(#1,#2)#3{\rlap{\L*=.1\Lengthunit
\xL*=#1\L* \yL*=#2\L*
\rx* \the\cos*\xL* \tmp* \the\sin*\yL* \advance\rx*-\tmp*
\ry* \the\cos*\yL* \tmp* \the\sin*\xL* \advance\ry*\tmp*
\kern\rx*\raise\ry*\hbox{\lower2pt\clap{$#3$}}}}

\def\sh*(#1,#2)#3{\rlap{\dm*=\the\n*\d**
\xL*=\xscale\dm* \yL*=\yscale\dm* \xL*=#1\xL* \yL*=#2\yL*
\rx* \the\cos*\xL* \tmp* \the\sin*\yL* \advance\rx*-\tmp*
\ry* \the\cos*\yL* \tmp* \the\sin*\xL* \advance\ry*\tmp*
\kern\rx*\raise\ry*\hbox{#3}}}

\def\calcnum*#1(#2,#3){\a*=1000sp\b*=1000sp\a*=#2\a*\b*=#3\b*
\ifdim\a*<0pt\a*-\a*\fi\ifdim\b*<0pt\b*-\b*\fi
\ifdim\a*>\b*\c*=.96\a*\advance\c*.4\b*
\else\c*=.96\b*\advance\c*.4\a*\fi
\k*\a*\multiply\k*\k*\l*\b*\multiply\l*\l*
\m*\k*\advance\m*\l*\n*\c*\r*\n*\multiply\n*\n*
\dn*\m*\advance\dn*-\n*\divide\dn*2\divide\dn*\r*
\advance\r*\dn*
\c*=\the\Nhalfperiods5sp\c*=#1\c*\ifdim\c*<0pt\c*-\c*\fi
\multiply\c*\r*\N*\c*\divide\N*10000}

\def\dashlin#1(#2,#3){\rlap{\calcnum*#1(#2,#3)\relax
\d**=#1\Lengthunit\ifdim\d**<0pt\d**-\d**\fi
\divide\N*2\multiply\N*2\advance\N*\*one
\divide\d**\N*\sm*\n*\*one\sh*(#2,#3){\sl*}\loop
\advance\n*\*one\sh*(#2,#3){\sm*}\advance\n*\*one
\sh*(#2,#3){\sl*}\ifnum\n*<\N*\repeat}}

\def\dashdotlin#1(#2,#3){\rlap{\calcnum*#1(#2,#3)\relax
\d**=#1\Lengthunit\ifdim\d**<0pt\d**-\d**\fi
\divide\N*2\multiply\N*2\advance\N*1\multiply\N*2\relax
\divide\d**\N*\sm*\n*\*two\sh*(#2,#3){\sl*}\loop
\advance\n*\*one\sh*(#2,#3){\kern-1.48pt\lower.5pt\hbox{\rm.}}\relax
\advance\n*\*one\sh*(#2,#3){\sm*}\advance\n*\*two
\sh*(#2,#3){\sl*}\ifnum\n*<\N*\repeat}}

\def\shl*(#1,#2)#3{\kern#1#3\lower#2#3\hbox{\unhcopy\spl*}}

\def\trianglin#1(#2,#3){\rlap{\toks0={#2}\toks1={#3}\calcnum*#1(#2,#3)\relax
\dd*=.57\Lengthunit\dd*=#1\dd*\divide\dd*\N*
\divide\dd*\*ths \multiply\dd*\magnitude
\d**=#1\Lengthunit\ifdim\d**<0pt\d**-\d**\fi
\multiply\N*2\divide\d**\N*\sm*\n*\*one\loop
\shl**{\dd*}\dd*-\dd*\advance\n*2\relax
\ifnum\n*<\N*\repeat\n*\N*\shl**{0pt}}}

\def\wavelin#1(#2,#3){\rlap{\toks0={#2}\toks1={#3}\calcnum*#1(#2,#3)\relax
\dd*=.23\Lengthunit\dd*=#1\dd*\divide\dd*\N*
\divide\dd*\*ths \multiply\dd*\magnitude
\d**=#1\Lengthunit\ifdim\d**<0pt\d**-\d**\fi
\multiply\N*4\divide\d**\N*\sm*\n*\*one\loop
\shl**{\dd*}\dt*=1.3\dd*\advance\n*\*one
\shl**{\dt*}\advance\n*\*one
\shl**{\dd*}\advance\n*\*two
\dd*-\dd*\ifnum\n*<\N*\repeat\n*\N*\shl**{0pt}}}

\def\w*lin(#1,#2){\rlap{\toks0={#1}\toks1={#2}\d**=\Lengthunit\dd*=-.12\d**
\divide\dd*\*ths \multiply\dd*\magnitude
\N*8\divide\d**\N*\sm*\n*\*one\loop
\shl**{\dd*}\dt*=1.3\dd*\advance\n*\*one
\shl**{\dt*}\advance\n*\*one
\shl**{\dd*}\advance\n*\*one
\shl**{0pt}\dd*-\dd*\advance\n*1\ifnum\n*<\N*\repeat}}

\def\l*arc(#1,#2)[#3][#4]{\rlap{\toks0={#1}\toks1={#2}\d**=\Lengthunit
\dd*=#3.037\d**\dd*=#4\dd*\dt*=#3.049\d**\dt*=#4\dt*\ifdim\d**>10mm\relax
\d**=.25\d**\n*\*one\shl**{-\dd*}\n*\*two\shl**{-\dt*}\n*3\relax
\shl**{-\dd*}\n*4\relax\shl**{0pt}\else
\ifdim\d**>5mm\d**=.5\d**\n*\*one\shl**{-\dt*}\n*\*two
\shl**{0pt}\else\n*\*one\shl**{0pt}\fi\fi}}

\def\d*arc(#1,#2)[#3][#4]{\rlap{\toks0={#1}\toks1={#2}\d**=\Lengthunit
\dd*=#3.037\d**\dd*=#4\dd*\d**=.25\d**\sm*\n*\*one\shl**{-\dd*}\relax
\n*3\relax\sh*(#1,#2){\xL*=\xscale\dd*\yL*=\yscale\dd*
\kern#2\xL*\lower#1\yL*\hbox{\sm*}}\n*4\relax\shl**{0pt}}}

\def\shl**#1{\c*=\the\n*\d**\d*=#1\relax
\a*=\the\toks0\c*\b*=\the\toks1\d*\advance\a*-\b*
\b*=\the\toks1\c*\d*=\the\toks0\d*\advance\b*\d*
\a*=\xscale\a*\b*=\yscale\b*
\rx* \the\cos*\a* \tmp* \the\sin*\b* \advance\rx*-\tmp*
\ry* \the\cos*\b* \tmp* \the\sin*\a* \advance\ry*\tmp*
\raise\ry*\rlap{\kern\rx*\unhcopy\spl*}}

\def\wlin*#1(#2,#3)[#4]{\rlap{\toks0={#2}\toks1={#3}\relax
\c*=#1\l*\c*\c*=.01\Lengthunit\m*\c*\divide\l*\m*
\c*=\the\Nhalfperiods5sp\multiply\c*\l*\N*\c*\divide\N*\*ths
\divide\N*2\multiply\N*2\advance\N*\*one
\dd*=.002\Lengthunit\dd*=#4\dd*\multiply\dd*\l*\divide\dd*\N*
\divide\dd*\*ths \multiply\dd*\magnitude
\d**=#1\multiply\N*4\divide\d**\N*\sm*\n*\*one\loop
\shl**{\dd*}\dt*=1.3\dd*\advance\n*\*one
\shl**{\dt*}\advance\n*\*one
\shl**{\dd*}\advance\n*\*two
\dd*-\dd*\ifnum\n*<\N*\repeat\n*\N*\shl**{0pt}}}

\def\wavebox#1{\setbox0\hbox{#1}\relax
\a*=\wd0\advance\a*14pt\b*=\ht0\advance\b*\dp0\advance\b*14pt\relax
\hbox{\kern9pt\relax
\rmov*(0pt,\ht0){\rmov*(-7pt,7pt){\wlin*\a*(1,0)[+]\wlin*\b*(0,-1)[-]}}\relax
\rmov*(\wd0,-\dp0){\rmov*(7pt,-7pt){\wlin*\a*(-1,0)[+]\wlin*\b*(0,1)[-]}}\relax
\box0\kern9pt}}

\def\rectangle#1(#2,#3){\relax
\lin#1(#2,0)\lin#1(0,#3)\mov#1(0,#3){\lin#1(#2,0)}\mov#1(#2,0){\lin#1(0,#3)}}

\def\dashrectangle#1(#2,#3){\dashlin#1(#2,0)\dashlin#1(0,#3)\relax
\mov#1(0,#3){\dashlin#1(#2,0)}\mov#1(#2,0){\dashlin#1(0,#3)}}

\def\waverectangle#1(#2,#3){\L*=#1\Lengthunit\a*=#2\L*\b*=#3\L*
\ifdim\a*<0pt\a*-\a*\def\x*{-1}\else\def\x*{1}\fi
\ifdim\b*<0pt\b*-\b*\def\y*{-1}\else\def\y*{1}\fi
\wlin*\a*(\x*,0)[-]\wlin*\b*(0,\y*)[+]\relax
\mov#1(0,#3){\wlin*\a*(\x*,0)[+]}\mov#1(#2,0){\wlin*\b*(0,\y*)[-]}}

\def\calcparab*{\ifnum\n*>\m*\k*\N*\advance\k*-\n*\else\k*\n*\fi
\a*=\the\k* sp\a*=10\a*\b*\dm*\advance\b*-\a*\k*\b*
\a*=\the\*ths\b*\divide\a*\l*\multiply\a*\k*
\divide\a*\l*\k*\*ths\r*\a*\advance\k*-\r*\dt*=\the\k*\L*}

\def\arcto#1(#2,#3)[#4]{\rlap{\toks0={#2}\toks1={#3}\calcnum*#1(#2,#3)\relax
\dm*=135sp\dm*=#1\dm*\d**=#1\Lengthunit\ifdim\dm*<0pt\dm*-\dm*\fi
\multiply\dm*\r*\a*=.3\dm*\a*=#4\a*\ifdim\a*<0pt\a*-\a*\fi
\advance\dm*\a*\N*\dm*\divide\N*10000\relax
\divide\N*2\multiply\N*2\advance\N*\*one
\L*=-.25\d**\L*=#4\L*\divide\d**\N*\divide\L*\*ths
\m*\N*\divide\m*2\dm*=\the\m*5sp\l*\dm*\sm*\n*\*one\loop
\calcparab*\shl**{-\dt*}\advance\n*1\ifnum\n*<\N*\repeat}}

\def\arrarcto#1(#2,#3)[#4]{\L*=#1\Lengthunit\L*=.54\L*
\arcto#1(#2,#3)[#4]\rmov*(#2\L*,#3\L*){\d*=.457\L*\d*=#4\d*\d**-\d*
\rmov*(#3\d**,#2\d*){\arrow.02(#2,#3)}}}

\def\dasharcto#1(#2,#3)[#4]{\rlap{\toks0={#2}\toks1={#3}\relax
\calcnum*#1(#2,#3)\dm*=\the\N*5sp\a*=.3\dm*\a*=#4\a*\ifdim\a*<0pt\a*-\a*\fi
\advance\dm*\a*\N*\dm*
\divide\N*20\multiply\N*2\advance\N*1\d**=#1\Lengthunit
\L*=-.25\d**\L*=#4\L*\divide\d**\N*\divide\L*\*ths
\m*\N*\divide\m*2\dm*=\the\m*5sp\l*\dm*
\sm*\n*\*one\loop\calcparab*
\shl**{-\dt*}\advance\n*1\ifnum\n*>\N*\else\calcparab*
\sh*(#2,#3){\xL*=#3\dt* \yL*=#2\dt*
\rx* \the\cos*\xL* \tmp* \the\sin*\yL* \advance\rx*\tmp*
\ry* \the\cos*\yL* \tmp* \the\sin*\xL* \advance\ry*-\tmp*
\kern\rx*\lower\ry*\hbox{\sm*}}\fi
\advance\n*1\ifnum\n*<\N*\repeat}}

\def\*shl*#1{\c*=\the\n*\d**\advance\c*#1\a**\d*\dt*\advance\d*#1\b**
\a*=\the\toks0\c*\b*=\the\toks1\d*\advance\a*-\b*
\b*=\the\toks1\c*\d*=\the\toks0\d*\advance\b*\d*
\rx* \the\cos*\a* \tmp* \the\sin*\b* \advance\rx*-\tmp*
\ry* \the\cos*\b* \tmp* \the\sin*\a* \advance\ry*\tmp*
\raise\ry*\rlap{\kern\rx*\unhcopy\spl*}}

\def\calcnormal*#1{\b**=10000sp\a**\b**\k*\n*\advance\k*-\m*
\multiply\a**\k*\divide\a**\m*\a**=#1\a**\ifdim\a**<0pt\a**-\a**\fi
\ifdim\a**>\b**\d*=.96\a**\advance\d*.4\b**
\else\d*=.96\b**\advance\d*.4\a**\fi
\d*=.01\d*\r*\d*\divide\a**\r*\divide\b**\r*
\ifnum\k*<0\a**-\a**\fi\d*=#1\d*\ifdim\d*<0pt\b**-\b**\fi
\k*\a**\a**=\the\k*\dd*\k*\b**\b**=\the\k*\dd*}

\def\wavearcto#1(#2,#3)[#4]{\rlap{\toks0={#2}\toks1={#3}\relax
\calcnum*#1(#2,#3)\c*=\the\N*5sp\a*=.4\c*\a*=#4\a*\ifdim\a*<0pt\a*-\a*\fi
\advance\c*\a*\N*\c*\divide\N*20\multiply\N*2\advance\N*-1\multiply\N*4\relax
\d**=#1\Lengthunit\dd*=.012\d**
\divide\dd*\*ths \multiply\dd*\magnitude
\ifdim\d**<0pt\d**-\d**\fi\L*=.25\d**
\divide\d**\N*\divide\dd*\N*\L*=#4\L*\divide\L*\*ths
\m*\N*\divide\m*2\dm*=\the\m*0sp\l*\dm*
\sm*\n*\*one\loop\calcnormal*{#4}\calcparab*
\*shl*{1}\advance\n*\*one\calcparab*
\*shl*{1.3}\advance\n*\*one\calcparab*
\*shl*{1}\advance\n*2\dd*-\dd*\ifnum\n*<\N*\repeat\n*\N*\shl**{0pt}}}

\def\triangarcto#1(#2,#3)[#4]{\rlap{\toks0={#2}\toks1={#3}\relax
\calcnum*#1(#2,#3)\c*=\the\N*5sp\a*=.4\c*\a*=#4\a*\ifdim\a*<0pt\a*-\a*\fi
\advance\c*\a*\N*\c*\divide\N*20\multiply\N*2\advance\N*-1\multiply\N*2\relax
\d**=#1\Lengthunit\dd*=.012\d**
\divide\dd*\*ths \multiply\dd*\magnitude
\ifdim\d**<0pt\d**-\d**\fi\L*=.25\d**
\divide\d**\N*\divide\dd*\N*\L*=#4\L*\divide\L*\*ths
\m*\N*\divide\m*2\dm*=\the\m*0sp\l*\dm*
\sm*\n*\*one\loop\calcnormal*{#4}\calcparab*
\*shl*{1}\advance\n*2\dd*-\dd*\ifnum\n*<\N*\repeat\n*\N*\shl**{0pt}}}

\def\hr*#1{\L*=\xscale\Lengthunit\ifnum
\angle**=0\clap{\vrule width#1\L* height.1pt}\else
\L*=#1\L*\L*=.5\L*\rmov*(-\L*,0pt){\sm*}\rmov*(\L*,0pt){\sl*}\fi}

\def\shade#1[#2]{\rlap{\Lengthunit=#1\Lengthunit
\special{em:linewidth .001pt}\relax
\mov(0,#2.05){\hr*{.994}}\mov(0,#2.1){\hr*{.980}}\relax
\mov(0,#2.15){\hr*{.953}}\mov(0,#2.2){\hr*{.916}}\relax
\mov(0,#2.25){\hr*{.867}}\mov(0,#2.3){\hr*{.798}}\relax
\mov(0,#2.35){\hr*{.715}}\mov(0,#2.4){\hr*{.603}}\relax
\mov(0,#2.45){\hr*{.435}}\special{em:linewidth \the\linwid*}}}

\def\dshade#1[#2]{\rlap{\special{em:linewidth .001pt}\relax
\Lengthunit=#1\Lengthunit\if#2-\def\t*{+}\else\def\t*{-}\fi
\mov(0,\t*.025){\relax
\mov(0,#2.05){\hr*{.995}}\mov(0,#2.1){\hr*{.988}}\relax
\mov(0,#2.15){\hr*{.969}}\mov(0,#2.2){\hr*{.937}}\relax
\mov(0,#2.25){\hr*{.893}}\mov(0,#2.3){\hr*{.836}}\relax
\mov(0,#2.35){\hr*{.760}}\mov(0,#2.4){\hr*{.662}}\relax
\mov(0,#2.45){\hr*{.531}}\mov(0,#2.5){\hr*{.320}}\relax
\special{em:linewidth \the\linwid*}}}}

\def\vdot{\rlap{\kern-1.9pt\lower1.8pt\hbox{$\scriptstyle\bullet$}}}
\def\vtimes{\rlap{\kern-3pt\lower1.8pt\hbox{$\scriptstyle\times$}}}
\def\vDot{\rlap{\kern-2.3pt\lower2.7pt\hbox{$\bullet$}}}
\def\vTimes{\rlap{\kern-3.6pt\lower2.4pt\hbox{$\times$}}}

\def\arc(#1)[#2,#3]{{\k*=#2\l*=#3\m*=\l*
\advance\m*-6\ifnum\k*>\l*\relax\else
{\rotate(#2)\mov(#1,0){\sm*}}\loop
\ifnum\k*<\m*\advance\k*5{\rotate(\k*)\mov(#1,0){\sl*}}\repeat
{\rotate(#3)\mov(#1,0){\sl*}}\fi}}

\def\dasharc(#1)[#2,#3]{{\k**=#2\n*=#3\advance\n*-1\advance\n*-\k**
\L*=1000sp\L*#1\L* \multiply\L*\n* \multiply\L*\Nhalfperiods
\divide\L*57\N*\L* \divide\N*2000\ifnum\N*=0\N*1\fi
\r*\n*  \divide\r*\N* \ifnum\r*<2\r*2\fi
\m**\r* \divide\m**2 \l**\r* \advance\l**-\m** \N*\n* \divide\N*\r*
\k**\r* \multiply\k**\N* \dn*\n* \advance\dn*-\k** \divide\dn*2\advance\dn*\*one
\r*\l** \divide\r*2\advance\dn*\r* \advance\N*-2\k**#2\relax
\ifnum\l**<6{\rotate(#2)\mov(#1,0){\sm*}}\advance\k**\dn*
{\rotate(\k**)\mov(#1,0){\sl*}}\advance\k**\m**
{\rotate(\k**)\mov(#1,0){\sm*}}\loop
\advance\k**\l**{\rotate(\k**)\mov(#1,0){\sl*}}\advance\k**\m**
{\rotate(\k**)\mov(#1,0){\sm*}}\advance\N*-1\ifnum\N*>0\repeat
{\rotate(#3)\mov(#1,0){\sl*}}\else\advance\k**\dn*
\arc(#1)[#2,\k**]\loop\advance\k**\m** \r*\k**
\advance\k**\l** {\arc(#1)[\r*,\k**]}\relax
\advance\N*-1\ifnum\N*>0\repeat
\advance\k**\m**\arc(#1)[\k**,#3]\fi}}

\def\triangarc#1(#2)[#3,#4]{{\k**=#3\n*=#4\advance\n*-\k**
\L*=1000sp\L*#2\L* \multiply\L*\n* \multiply\L*\Nhalfperiods
\divide\L*57\N*\L* \divide\N*1000\ifnum\N*=0\N*1\fi
\d**=#2\Lengthunit \d*\d** \divide\d*57\multiply\d*\n*
\r*\n*  \divide\r*\N* \ifnum\r*<2\r*2\fi
\m**\r* \divide\m**2 \l**\r* \advance\l**-\m** \N*\n* \divide\N*\r*
\dt*\d* \divide\dt*\N* \dt*.5\dt* \dt*#1\dt*
\divide\dt*1000\multiply\dt*\magnitude
\k**\r* \multiply\k**\N* \dn*\n* \advance\dn*-\k** \divide\dn*2\relax
\r*\l** \divide\r*2\advance\dn*\r* \advance\N*-1\k**#3\relax
{\rotate(#3)\mov(#2,0){\sm*}}\advance\k**\dn*
{\rotate(\k**)\mov(#2,0){\sl*}}\advance\k**-\m**\advance\l**\m**\loop\dt*-\dt*
\d*\d** \advance\d*\dt*
\advance\k**\l**{\rotate(\k**)\rmov*(\d*,0pt){\sl*}}%
\advance\N*-1\ifnum\N*>0\repeat\advance\k**\m**
{\rotate(\k**)\mov(#2,0){\sl*}}{\rotate(#4)\mov(#2,0){\sl*}}}}

\def\wavearc#1(#2)[#3,#4]{{\k**=#3\n*=#4\advance\n*-\k**
\L*=4000sp\L*#2\L* \multiply\L*\n* \multiply\L*\Nhalfperiods
\divide\L*57\N*\L* \divide\N*1000\ifnum\N*=0\N*1\fi
\d**=#2\Lengthunit \d*\d** \divide\d*57\multiply\d*\n*
\r*\n*  \divide\r*\N* \ifnum\r*=0\r*1\fi
\m**\r* \divide\m**2 \l**\r* \advance\l**-\m** \N*\n* \divide\N*\r*
\dt*\d* \divide\dt*\N* \dt*.7\dt* \dt*#1\dt*
\divide\dt*1000\multiply\dt*\magnitude
\k**\r* \multiply\k**\N* \dn*\n* \advance\dn*-\k** \divide\dn*2\relax
\divide\N*4\advance\N*-1\k**#3\relax
{\rotate(#3)\mov(#2,0){\sm*}}\advance\k**\dn*
{\rotate(\k**)\mov(#2,0){\sl*}}\advance\k**-\m**\advance\l**\m**\loop\dt*-\dt*
\d*\d** \advance\d*\dt* \dd*\d** \advance\dd*1.3\dt*
\advance\k**\r*{\rotate(\k**)\rmov*(\d*,0pt){\sl*}}\relax
\advance\k**\r*{\rotate(\k**)\rmov*(\dd*,0pt){\sl*}}\relax
\advance\k**\r*{\rotate(\k**)\rmov*(\d*,0pt){\sl*}}\relax
\advance\k**\r*
\advance\N*-1\ifnum\N*>0\repeat\advance\k**\m**
{\rotate(\k**)\mov(#2,0){\sl*}}{\rotate(#4)\mov(#2,0){\sl*}}}}

\def\gmov*#1(#2,#3)#4{\rlap{\L*=#1\Lengthunit
\xL*=#2\L* \yL*=#3\L*
\rx* \gcos*\xL* \tmp* \gsin*\yL* \advance\rx*-\tmp*
\ry* \gcos*\yL* \tmp* \gsin*\xL* \advance\ry*\tmp*
\rx*=\xscale\rx* \ry*=\yscale\ry*
\xL* \the\cos*\rx* \tmp* \the\sin*\ry* \advance\xL*-\tmp*
\yL* \the\cos*\ry* \tmp* \the\sin*\rx* \advance\yL*\tmp*
\kern\xL*\raise\yL*\hbox{#4}}}

\def\rgmov*(#1,#2)#3{\rlap{\xL*#1\yL*#2\relax
\rx* \gcos*\xL* \tmp* \gsin*\yL* \advance\rx*-\tmp*
\ry* \gcos*\yL* \tmp* \gsin*\xL* \advance\ry*\tmp*
\rx*=\xscale\rx* \ry*=\yscale\ry*
\xL* \the\cos*\rx* \tmp* \the\sin*\ry* \advance\xL*-\tmp*
\yL* \the\cos*\ry* \tmp* \the\sin*\rx* \advance\yL*\tmp*
\kern\xL*\raise\yL*\hbox{#3}}}

\def\Earc(#1)[#2,#3][#4,#5]{{\k*=#2\l*=#3\m*=\l*
\advance\m*-6\ifnum\k*>\l*\relax\else\def\xscale{#4}\def\yscale{#5}\relax
{\angle**0\rotate(#2)}\gmov*(#1,0){\sm*}\loop
\ifnum\k*<\m*\advance\k*5\relax
{\angle**0\rotate(\k*)}\gmov*(#1,0){\sl*}\repeat
{\angle**0\rotate(#3)}\gmov*(#1,0){\sl*}\relax
\def\xscale{1}\def\yscale{1}\fi}}

\def\dashEarc(#1)[#2,#3][#4,#5]{{\k**=#2\n*=#3\advance\n*-1\advance\n*-\k**
\L*=1000sp\L*#1\L* \multiply\L*\n* \multiply\L*\Nhalfperiods
\divide\L*57\N*\L* \divide\N*2000\ifnum\N*=0\N*1\fi
\r*\n*  \divide\r*\N* \ifnum\r*<2\r*2\fi
\m**\r* \divide\m**2 \l**\r* \advance\l**-\m** \N*\n* \divide\N*\r*
\k**\r*\multiply\k**\N* \dn*\n* \advance\dn*-\k** \divide\dn*2\advance\dn*\*one
\r*\l** \divide\r*2\advance\dn*\r* \advance\N*-2\k**#2\relax
\ifnum\l**<6\def\xscale{#4}\def\yscale{#5}\relax
{\angle**0\rotate(#2)}\gmov*(#1,0){\sm*}\advance\k**\dn*
{\angle**0\rotate(\k**)}\gmov*(#1,0){\sl*}\advance\k**\m**
{\angle**0\rotate(\k**)}\gmov*(#1,0){\sm*}\loop
\advance\k**\l**{\angle**0\rotate(\k**)}\gmov*(#1,0){\sl*}\advance\k**\m**
{\angle**0\rotate(\k**)}\gmov*(#1,0){\sm*}\advance\N*-1\ifnum\N*>0\repeat
{\angle**0\rotate(#3)}\gmov*(#1,0){\sl*}\def\xscale{1}\def\yscale{1}\else
\advance\k**\dn* \Earc(#1)[#2,\k**][#4,#5]\loop\advance\k**\m** \r*\k**
\advance\k**\l** {\Earc(#1)[\r*,\k**][#4,#5]}\relax
\advance\N*-1\ifnum\N*>0\repeat
\advance\k**\m**\Earc(#1)[\k**,#3][#4,#5]\fi}}

\def\triangEarc#1(#2)[#3,#4][#5,#6]{{\k**=#3\n*=#4\advance\n*-\k**
\L*=1000sp\L*#2\L* \multiply\L*\n* \multiply\L*\Nhalfperiods
\divide\L*57\N*\L* \divide\N*1000\ifnum\N*=0\N*1\fi
\d**=#2\Lengthunit \d*\d** \divide\d*57\multiply\d*\n*
\r*\n*  \divide\r*\N* \ifnum\r*<2\r*2\fi
\m**\r* \divide\m**2 \l**\r* \advance\l**-\m** \N*\n* \divide\N*\r*
\dt*\d* \divide\dt*\N* \dt*.5\dt* \dt*#1\dt*
\divide\dt*1000\multiply\dt*\magnitude
\k**\r* \multiply\k**\N* \dn*\n* \advance\dn*-\k** \divide\dn*2\relax
\r*\l** \divide\r*2\advance\dn*\r* \advance\N*-1\k**#3\relax
\def\xscale{#5}\def\yscale{#6}\relax
{\angle**0\rotate(#3)}\gmov*(#2,0){\sm*}\advance\k**\dn*
{\angle**0\rotate(\k**)}\gmov*(#2,0){\sl*}\advance\k**-\m**
\advance\l**\m**\loop\dt*-\dt* \d*\d** \advance\d*\dt*
\advance\k**\l**{\angle**0\rotate(\k**)}\rgmov*(\d*,0pt){\sl*}\relax
\advance\N*-1\ifnum\N*>0\repeat\advance\k**\m**
{\angle**0\rotate(\k**)}\gmov*(#2,0){\sl*}\relax
{\angle**0\rotate(#4)}\gmov*(#2,0){\sl*}\def\xscale{1}\def\yscale{1}}}

\def\waveEarc#1(#2)[#3,#4][#5,#6]{{\k**=#3\n*=#4\advance\n*-\k**
\L*=4000sp\L*#2\L* \multiply\L*\n* \multiply\L*\Nhalfperiods
\divide\L*57\N*\L* \divide\N*1000\ifnum\N*=0\N*1\fi
\d**=#2\Lengthunit \d*\d** \divide\d*57\multiply\d*\n*
\r*\n*  \divide\r*\N* \ifnum\r*=0\r*1\fi
\m**\r* \divide\m**2 \l**\r* \advance\l**-\m** \N*\n* \divide\N*\r*
\dt*\d* \divide\dt*\N* \dt*.7\dt* \dt*#1\dt*
\divide\dt*1000\multiply\dt*\magnitude
\k**\r* \multiply\k**\N* \dn*\n* \advance\dn*-\k** \divide\dn*2\relax
\divide\N*4\advance\N*-1\k**#3\def\xscale{#5}\def\yscale{#6}\relax
{\angle**0\rotate(#3)}\gmov*(#2,0){\sm*}\advance\k**\dn*
{\angle**0\rotate(\k**)}\gmov*(#2,0){\sl*}\advance\k**-\m**
\advance\l**\m**\loop\dt*-\dt*
\d*\d** \advance\d*\dt* \dd*\d** \advance\dd*1.3\dt*
\advance\k**\r*{\angle**0\rotate(\k**)}\rgmov*(\d*,0pt){\sl*}\relax
\advance\k**\r*{\angle**0\rotate(\k**)}\rgmov*(\dd*,0pt){\sl*}\relax
\advance\k**\r*{\angle**0\rotate(\k**)}\rgmov*(\d*,0pt){\sl*}\relax
\advance\k**\r*
\advance\N*-1\ifnum\N*>0\repeat\advance\k**\m**
{\angle**0\rotate(\k**)}\gmov*(#2,0){\sl*}\relax
{\angle**0\rotate(#4)}\gmov*(#2,0){\sl*}\def\xscale{1}\def\yscale{1}}}

\newcount\CatcodeOfAtSign
\CatcodeOfAtSign=\the\catcode`\@
\catcode`\@=11
\def\@arc#1[#2][#3]{\rlap{\Lengthunit=#1\Lengthunit
\sm*\l*arc(#2.1914,#3.0381)[#2][#3]\relax
\mov(#2.1914,#3.0381){\l*arc(#2.1622,#3.1084)[#2][#3]}\relax
\mov(#2.3536,#3.1465){\l*arc(#2.1084,#3.1622)[#2][#3]}\relax
\mov(#2.4619,#3.3086){\l*arc(#2.0381,#3.1914)[#2][#3]}}}

\def\dash@arc#1[#2][#3]{\rlap{\Lengthunit=#1\Lengthunit
\d*arc(#2.1914,#3.0381)[#2][#3]\relax
\mov(#2.1914,#3.0381){\d*arc(#2.1622,#3.1084)[#2][#3]}\relax
\mov(#2.3536,#3.1465){\d*arc(#2.1084,#3.1622)[#2][#3]}\relax
\mov(#2.4619,#3.3086){\d*arc(#2.0381,#3.1914)[#2][#3]}}}

\def\wave@arc#1[#2][#3]{\rlap{\Lengthunit=#1\Lengthunit
\w*lin(#2.1914,#3.0381)\relax
\mov(#2.1914,#3.0381){\w*lin(#2.1622,#3.1084)}\relax
\mov(#2.3536,#3.1465){\w*lin(#2.1084,#3.1622)}\relax
\mov(#2.4619,#3.3086){\w*lin(#2.0381,#3.1914)}}}

\def\bezier#1(#2,#3)(#4,#5)(#6,#7){\N*#1\l*\N* \advance\l*\*one
\d* #4\Lengthunit \advance\d* -#2\Lengthunit \multiply\d* \*two
\b* #6\Lengthunit \advance\b* -#2\Lengthunit
\advance\b*-\d* \divide\b*\N*
\d** #5\Lengthunit \advance\d** -#3\Lengthunit \multiply\d** \*two
\b** #7\Lengthunit \advance\b** -#3\Lengthunit
\advance\b** -\d** \divide\b**\N*
\mov(#2,#3){\sm*{\loop\ifnum\m*<\l*
\a*\m*\b* \advance\a*\d* \divide\a*\N* \multiply\a*\m*
\a**\m*\b** \advance\a**\d** \divide\a**\N* \multiply\a**\m*
\rmov*(\a*,\a**){\unhcopy\spl*}\advance\m*\*one\repeat}}}

\catcode`\*=12

\newcount\n@ast
\def\n@ast@#1{\n@ast0\relax\get@ast@#1\end}
\def\get@ast@#1{\ifx#1\end\let\next\relax\else
\ifx#1*\advance\n@ast1\fi\let\next\get@ast@\fi\next}

\newif\if@up \newif\if@dwn
\def\up@down@#1{\@upfalse\@dwnfalse
\if#1u\@uptrue\fi\if#1U\@uptrue\fi\if#1+\@uptrue\fi
\if#1d\@dwntrue\fi\if#1D\@dwntrue\fi\if#1-\@dwntrue\fi}

\def\halfcirc#1(#2)[#3]{{\Lengthunit=#2\Lengthunit\up@down@{#3}\relax
\if@up\mov(0,.5){\@arc[-][-]\@arc[+][-]}\fi
\if@dwn\mov(0,-.5){\@arc[-][+]\@arc[+][+]}\fi
\def\lft{\mov(0,.5){\@arc[-][-]}\mov(0,-.5){\@arc[-][+]}}\relax
\def\rght{\mov(0,.5){\@arc[+][-]}\mov(0,-.5){\@arc[+][+]}}\relax
\if#3l\lft\fi\if#3L\lft\fi\if#3r\rght\fi\if#3R\rght\fi
\n@ast@{#1}\relax
\ifnum\n@ast>0\if@up\shade[+]\fi\if@dwn\shade[-]\fi\fi
\ifnum\n@ast>1\if@up\dshade[+]\fi\if@dwn\dshade[-]\fi\fi}}

\def\halfdashcirc(#1)[#2]{{\Lengthunit=#1\Lengthunit\up@down@{#2}\relax
\if@up\mov(0,.5){\dash@arc[-][-]\dash@arc[+][-]}\fi
\if@dwn\mov(0,-.5){\dash@arc[-][+]\dash@arc[+][+]}\fi
\def\lft{\mov(0,.5){\dash@arc[-][-]}\mov(0,-.5){\dash@arc[-][+]}}\relax
\def\rght{\mov(0,.5){\dash@arc[+][-]}\mov(0,-.5){\dash@arc[+][+]}}\relax
\if#2l\lft\fi\if#2L\lft\fi\if#2r\rght\fi\if#2R\rght\fi}}

\def\halfwavecirc(#1)[#2]{{\Lengthunit=#1\Lengthunit\up@down@{#2}\relax
\if@up\mov(0,.5){\wave@arc[-][-]\wave@arc[+][-]}\fi
\if@dwn\mov(0,-.5){\wave@arc[-][+]\wave@arc[+][+]}\fi
\def\lft{\mov(0,.5){\wave@arc[-][-]}\mov(0,-.5){\wave@arc[-][+]}}\relax
\def\rght{\mov(0,.5){\wave@arc[+][-]}\mov(0,-.5){\wave@arc[+][+]}}\relax
\if#2l\lft\fi\if#2L\lft\fi\if#2r\rght\fi\if#2R\rght\fi}}

\catcode`\*=11

\def\Circle#1(#2){\halfcirc#1(#2)[u]\halfcirc#1(#2)[d]\n@ast@{#1}\relax
\ifnum\n@ast>0\L*=\xscale\Lengthunit
\ifnum\angle**=0\clap{\vrule width#2\L* height.1pt}\else
\L*=#2\L*\L*=.5\L*\special{em:linewidth .001pt}\relax
\rmov*(-\L*,0pt){\sm*}\rmov*(\L*,0pt){\sl*}\relax
\special{em:linewidth \the\linwid*}\fi\fi}

\catcode`\*=12

\def\wavecirc(#1){\halfwavecirc(#1)[u]\halfwavecirc(#1)[d]}

\def\dashcirc(#1){\halfdashcirc(#1)[u]\halfdashcirc(#1)[d]}

\def\xscale{1}
\def\yscale{1}

\def\Ellipse#1(#2)[#3,#4]{\def\xscale{#3}\def\yscale{#4}\relax
\Circle#1(#2)\def\xscale{1}\def\yscale{1}}

\def\dashEllipse(#1)[#2,#3]{\def\xscale{#2}\def\yscale{#3}\relax
\dashcirc(#1)\def\xscale{1}\def\yscale{1}}

\def\waveEllipse(#1)[#2,#3]{\def\xscale{#2}\def\yscale{#3}\relax
\wavecirc(#1)\def\xscale{1}\def\yscale{1}}

\def\halfEllipse#1(#2)[#3][#4,#5]{\def\xscale{#4}\def\yscale{#5}\relax
\halfcirc#1(#2)[#3]\def\xscale{1}\def\yscale{1}}

\def\halfdashEllipse(#1)[#2][#3,#4]{\def\xscale{#3}\def\yscale{#4}\relax
\halfdashcirc(#1)[#2]\def\xscale{1}\def\yscale{1}}

\def\halfwaveEllipse(#1)[#2][#3,#4]{\def\xscale{#3}\def\yscale{#4}\relax
\halfwavecirc(#1)[#2]\def\xscale{1}\def\yscale{1}}

\catcode`\@=\the\CatcodeOfAtSign

\section{Introduction}

Experimental and theoretical investigations of the g-factor of a free and
bound leptons (the electron and muon) are important both for the check of
the Standard Model and determination more correct values of several
fundamental parameters of the theory including the lepton-to-proton mass
ratio \cite{MT,TB}. The growth of the interest to theoretical
calculations on this problem in the last years is connected with new
experiments where the hydrogen-like ions with moderate values of the nuclear
charge $Z$ are used \cite{BT} (see experimental data shown in Table \ref{t1}).
An increase of the experimental accuracy
for the electron g-factor measurements in such simple atoms generates a need
for the calculation of new corrections which can become essential for the
comparison of the theory and experiment.

Gyromagnetic factor of the bound lepton can be represented in the form:
\begin{equation}
g_l=2+\Delta g_{rel}+\Delta g_{rad}+\Delta g_{rec}+\Delta g_{str}+...  .
\end{equation}
Relativistic corrections $\Delta g_{rel}$, radiative corrections
$\Delta g_{rad}$, recoil corrections $\Delta g_{rec}$ were investigated
in Refs.\cite{F1,F2,G,GH,CO,EG,BCS,PSSL,VS,K} with an accuracy
up to fourth order terms.
Certain contributions to the $g_l$ of higher order over $\alpha$ related
for the most part to the vacuum polarization and the lepton self energy
were studied in Refs.\cite{K,Y,KIS}. An agreement between theoretical and
experimental results in the problem of the gyromagnetic factor of the bound
electron is observed to the present.
For hydrogenic ions with sufficiently high values $Z$ the nuclear structure
effects $\Delta g_{str}$ acquire important role. They can be taken into
account by electromagnetic form factors of minimal multipolarity
describing the distributions of the electric charge and magnetic moment.
Main contribution of order $(Z\alpha)^4$ to the $\Delta g_{str}$ is
determined by the nuclear charge radius $R_N$ which is the differential parameter
of the electric charge distribution. It was originally obtained on the basis
of nonrelativistic approach in Ref.\cite{K} and by means of the Dirac
equation in Ref.\cite{GS} where relativistic correction of order
$(Z\alpha)^6m_1^2R^2_N$ to the $\Delta g_{str}$ was found also.
The quasipotential method for the problem of magnetic moment of hydrogenic
atom was formulated in Refs.\cite{F1,F2}. In this approach the two-particle
vertex function describing the interaction of the bound state with external
electromagnetic field was introduced. Its expansion by the perturbative
series for hydrogenic systems allows to calculate radiative, recoil corrections
of different orders. In the case of the constituent particles with arbitrary
spin the construction of the vertex operator in two main orders by the perturbative
series was carried out in Ref.\cite{MF1}. The aim of present work consists in
the calculation of the nuclear structure effects of orders $(Z\alpha)^4$ and
$(Z\alpha)^5$ including recoil effects in the g-factors of the bound electron
and muon on the basis of the quasipotential method. Considering that in recent
experiments with hydrogenic ions an accuracy comprises several ppb \cite{TR,WPK,LR,JR,H,Werth}
it is evident that the role of the nuclear structure corrections in the lepton
g-factor for moderate and high $Z$ will be enhance.

\begin{table}
\caption{\label{t1} Experimental data and theoretical values on the g-factors
of the bound electron.}
\bigskip
\begin{ruledtabular}
\begin{tabular}{|c|c|c|c|}   \hline
Ratio& Experiment& Ref.& Theory \\
of the g-factors         &            & & \cite{MT,BT}      \\    \hline
${\rm\frac{g_{e^-}(H)}{g_{e^-}}}$ & ${\rm 1-17.709(13)\times 10^{-6}}$&\cite{TR}&${\rm 1-17.7053\times 10^{-6}}$ \\   \hline
${\rm\frac{g_{e^-}(D)}{g_{e^-}}}$ & ---& ---&${\rm 1-17.7125\times 10^{-6}}$ \\   \hline
${\rm\frac{g_{e^-}(Mu)}{g_{e^-}}}$ & ---& ---&${\rm 1-17.591\times 10^{-6}}$ \\   \hline
${\rm\frac{g_{e^-}(H)}{g_{e^-}(D)}}$ & ${\rm 1+7.22(3)\times 10^{-9}}$& \cite{WPK}&${\rm 1+7.247\times 10^{-9}}$ \\   \hline
${\rm\frac{g_{e^-}(H)}{g_{e^-}(T)}}$ & ${\rm 1+10.7(1.5)\times 10^{-9}}$&\cite{LR}&${\rm 1+10.7\times 10^{-9}}$ \\   \hline
${\rm\frac{g_{e^-}(^4He^+)}{g_{e^-}}}$ & ${\rm 1-70.87(30)\times 10^{-6}}$&\cite{JR}&${\rm 1-70.87\times 10^{-6}}$ \\  \hline
${\rm\frac{g_{e^-}(^{12}C^{5+})}{2}}$ & ${\rm 1+520.798(2)\times 10^{-6}}$&\cite{H}&${\rm 1+520.796\times 10^{-6}}$ \\  \hline
${\rm\frac{g_{e^-}(^{16}O^{7+})}{2}}$ & ${\rm 1+23.514(2)\times 10^{-6}}$&\cite{Werth}&${\rm 1+23.511\times 10^{-6}}$ \\  \hline
\end{tabular}
\end{ruledtabular}
\end{table}

\section{General formalism}

Magnetic moment of the two-particle bound state can be written as follows
\cite{F1,F2}:
\begin{equation}
{\cal\bf M}=-\frac{i}{2}\left[\frac{\partial}{\partial\bf\Delta}\times
<\Psi_{n,Q}|{\bf J}(0)|\Psi_{n,P}>\right],~~{\bf\Delta}={\bf Q}-{\bf P},
\end{equation}
where the matrix element of electromagnetic current operator ${\bf J}$
between the bound states with total momenta ${\bf Q}$ and ${\bf P}$
respectively can be expressed in terms of the wave functions of the bound
system $\Psi_{n,Q}({\bf q})$, $\Psi_{n,P}({\bf p})$ and generalized vertex
function $\Gamma_\mu$ represented in Fig.1 in the form:
\begin{equation}
<\Psi_{n,Q}|J_\mu(0)|\Psi_{n,P}>=\int\frac{d{\bf p} d{\bf q}}{(2\pi)^6}
\bar\Psi_{n,Q}({\bf q})\Gamma_\mu({\bf p},{\bf q})
\Psi_{n,P}({\bf p}),
\end{equation}
The two-particle vertex function $\Gamma_\mu$ is determined in terms of the
five-point Green-like function
\begin{equation}
R_\mu=<0|\psi_1(t,{\bf x_1})\psi_2(t,{\bf x_2})J_\mu(0)\bar\psi_1(\tau,
{\bf y_1})\bar\psi_2(\tau,{\bf y_2})|0>,
\end{equation}
projected on the positive energy states \cite{F1,F2}:
\begin{equation}
\Gamma_\mu=G^{-1}R^{(+)}G^{-1},~~~R^{(+)}_\mu=\bar u_1\bar u_2R_\mu u_1u_2,
\end{equation}
where G is the two-particle Green function and $u_1$, $u_2$ are the wave functions
of a free particles. For loosely bound system in quantum electrodynamics all
introduced quantities $\Gamma$, $G^{-1}$, $R$ can be represented
by the perturbative series:
\begin{equation}
\Gamma=\Gamma^{(0)}+\Gamma^{(1)}+\Gamma^{(2)}+..., R=R_0+R_1+R_2+...,
G^{-1}=G^{-1}_0-V_1-...  .
\end{equation}
Substituting Eq.(6) in Eq.(5) we obtain the following relations
for generalized vertex function:
\begin{equation}
\Gamma^{(0)}=G^{-1}_0R_0G^{-1}_0, \Gamma^{(1)}=G^{-1}_0R_1G^{-1}_0-
V_1G_0\Gamma^{(0)}-\Gamma^{(0)}G_0V_1,
\end{equation}
\begin{displaymath}
\Gamma^{(2)}=G^{-1}_0R_2G^{-1}_0-V_2G_0\Gamma^{(0)}-\Gamma^{(0)}G_0V_2-
V_1G_0\Gamma^{(1)}-\Gamma^{(1)}G_0V_1-2V_1G_0\Gamma^{(0)}G_0V_1,
\end{displaymath}
where $G_0$ is the Green function of two noninteracting particles,
$V_1$ is the quasipotential of the one-photon interaction, $V_2$ is
the quasipotential of the two-photon interaction.

\begin{figure}
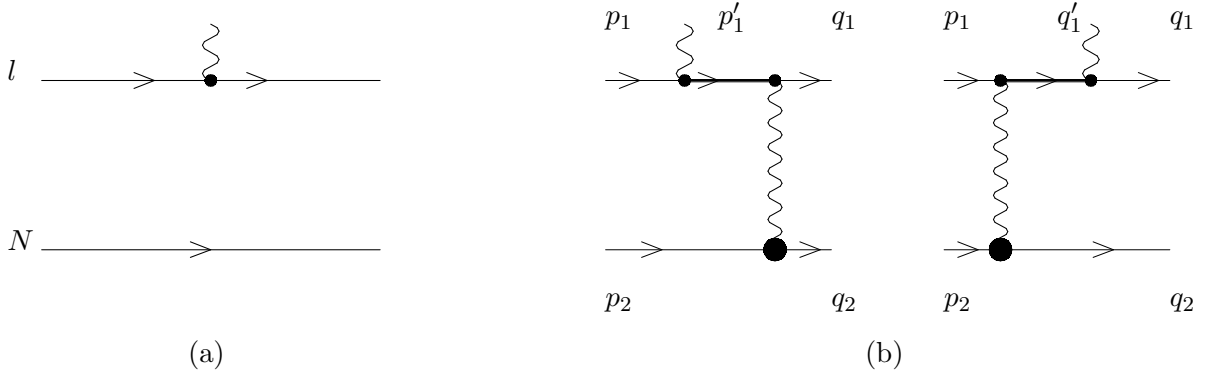

\magnitude=2000
\GRAPH(hsize=15){
\mov(6,0){\lin(2,0)}%
\mov(6.5,0){\lin(-0.18,-0.07)}%
\mov(6.5,0){\lin(-0.18,0.07)}%
\mov(6.3,1.5){\lin(-0.18,-0.07)}%
\mov(6.3,1.5){\lin(-0.18,0.07)}%
\mov(7.9,0){\lin(-0.18,-0.07)}%
\mov(7.9,0){\lin(-0.18,0.07)}%
\mov(7,1.5){\lin(-0.18,-0.07)}%
\mov(7,1.5){\lin(-0.18,0.07)}%
\mov(7.9,1.5){\lin(-0.18,-0.07)}%
\mov(7.9,1.5){\lin(-0.18,0.07)}%
\mov(9.3,0){\lin(-0.18,-0.07)}%
\mov(9.3,0){\lin(-0.18,0.07)}%
\mov(9.3,1.5){\lin(-0.18,-0.07)}%
\mov(9.3,1.5){\lin(-0.18,0.07)}%
\mov(10.5,0){\lin(-0.18,-0.07)}%
\mov(10.5,0){\lin(-0.18,0.07)}%
\mov(10,1.5){\lin(-0.18,-0.07)}%
\mov(10,1.5){\lin(-0.18,0.07)}%
\mov(10.9,1.5){\lin(-0.18,-0.07)}%
\mov(10.9,1.5){\lin(-0.18,0.07)}%
\mov(6,-0.5){$p_2$}%
\mov(6,2){$p_1$}%
\mov(8,-0.5){$q_2$}%
\mov(8,2.){$q_1$}%
\mov(7.,2){$p'_1$}%
\mov(9,-0.5){$p_2$}%
\mov(9,2){$p_1$}%
\mov(11,2){$q_1$}%
\mov(11,-0.5){$q_2$}%
\mov(10,2){$q'_1$}%
\mov(9,0){\lin(2,0)}%
\mov(6,1.5){\lin(2,0)}%
\mov(9,1.5){\lin(2,0)}%
\mov(6.7,1.5){\Circle*(0.1)}%
\mov(7.5,1.5){\wavelin(0,-1.5)}%
\mov(7.5,1.5){\Circle*(0.1)}%
\mov(7.5,1.51){\lin(-0.8,0)}%
\mov(7.5,1.49){\lin(-0.8,0)}%
\mov(9.5,1.5){\Circle*(0.1)}%
\mov(9.5,1.51){\lin(0.8,0)}%
\mov(9.5,1.49){\lin(0.8,0)}%
\mov(7.5,0.){\Circle**(0.2)}%
\mov(9.5,0.){\Circle**(0.2)}%
\mov(8.3,-1.){(b)}%
\mov(2.3,-1.){(a)}%
\mov(6.7,1.5){\wavelin(0,0.5)}%
\mov(9.5,1.5){\wavelin(0,-1.5)}%
\mov(10.3,1.5){\Circle*(0.1)}%
\mov(10.3,1.5){\wavelin(0,0.5)}%
\mov(1.,0){\lin(3,0)}%
\mov(0.7,0){$N$}%
\mov(0.7,1.5){$l$}%
\mov(1,1.5){\lin(3,0)}%
\mov(2.5,1.5){\Circle*(0.1)}%
\mov(2.5,1.5){\wavelin(0,0.5)}%
\mov(2.5,0){\lin(-0.18,-0.07)}%
\mov(2.5,0){\lin(-0.18,0.07)}%
\mov(2.,1.5){\lin(-0.18,-0.07)}%
\mov(2.,1.5){\lin(-0.18,0.07)}%
\mov(3.,1.5){\lin(-0.18,-0.07)}%
\mov(3.,1.5){\lin(-0.18,0.07)}%
}
\vspace{3mm}
\caption{Generalized two-particle vertex function $\Gamma_\mu$:
the diagram (a) represents $\Gamma_\mu^{(0)}$ with free noninteracting
particles, the diagrams (b)
determine the nucleus ($N$) structure corrections of order $(Z\alpha)^4$
in the g-factor of the bound lepton $(l)$.
Bold line denotes negative energy part of the lepton propagator.
Large bold circle denotes the interaction vertex of the nucleus with
electromagnetic field proportional to the factor $\left(F_1(k^2)-1\right)$.}
\end{figure}

The transformation law for the wave function $\Psi_{n,P}({\bf p})$ of two
bound particles with spins $s_1$, $s_2$ from the rest frame to the reference
frame moving with momentum ${\bf P}$ was obtained in Ref.\cite{F3} in the form:
\begin{equation}
\Psi_{n,P}({\bf p})=D^{s_1}_1(R^W_{L_P})D^{s_2}_2(R^W_{L_P})\Psi_{n,0}({\bf p}),
\end{equation}
where $D^s(R)$ is the rotation matrix, $R^W$ is the Wigner rotation and
$L_P$ is the Lorentz boost from the rest frame to the reference frame moving
with momentum ${\bf P}$. The rotation matrix can be expressed in terms of the
Lorentz transformation matrices as follows:
\begin{equation}
D^s(R^W_{L_P})=S^{-1}({\bf p}_{1,2})S({\bf P})S({\bf p}).
\end{equation}
The quasipotential bound state wave function $\Psi_0({\bf p})$ in the rest
frame of the composite system satisfies the quasipotential equation
\cite{MF2}:
\begin{equation}
G^{-1}_0\Psi\equiv\left(\frac{b^2}{2\mu_R}-\frac{{\bf p}^2}{2\mu_R}\right)
\Psi_0({\bf p})=\int V({\bf p},{\bf q},M)\Psi_0({\bf q})
\frac{d{\bf q}}{(2\pi)^3},
\end{equation}
where $\mu_R$ is relativistic reduced mass:
\begin{equation}
\mu_R=\frac{E_1E_2}{M}=\frac{M^4-(m_1^2-m_2^2)^2}{4M^3}, E_{1,2}=\frac{M^2-m_{2,1}^2+m_{1,2}^2}
{2M},
\end{equation}
$M=E_1+E_2$ is the mass of the bound state,
\begin{equation}
b^2(M)=\frac{[M^2-(m_1+m_2)^2][M^2-(m_1-m_2)^2]}{4M^2}.
\end{equation}
In the nonrelativistic limit the equation (10) reduces to the Shroedinger
equation with the Coulomb potential. The main contribution to the vertex
function $\Gamma_\mu$ is equal
\begin{equation}
\Gamma^{(0)\lambda\sigma,\rho\omega}_\mu({\bf p},{\bf q})=\bar
u_1^\lambda({\bf q_1})\gamma_\mu u_1^\rho({\bf p_1})\bar u_2^\sigma({\bf q_2})
u_2^\omega({\bf p_2})(2\pi)^3\delta({\bf p_2}-{\bf q_2})\delta^{\sigma\omega},
\lambda,\sigma,\rho,\omega=\pm\frac{1}{2},
\end{equation}
where the Dirac spinors are
\begin{equation}
u^\lambda({p})=\sqrt{\frac{\epsilon(p)+m}{2\epsilon(p)}}
\left(\begin{array}{c}
1\\ \displaystyle\frac{(\mathstrut\bm{\sigma}{\bf p})}{\epsilon(p)+m}
\end{array}\right)
\chi^\lambda, ~~\lambda=\pm\frac{1}{2}.
\end{equation}
Relativistic four momenta of the particles in the initial and final states
are defined as follows:
\begin{equation}
p_{1,2}=\varepsilon_{1,2}({\bf p})v\pm\sum_{i=1}^3n^{(i)}(v)p^i,~~~v=\frac{P}{M},
\end{equation}
\begin{displaymath}
q_{1,2}=\varepsilon_{1,2}({\bf q})v'\pm\sum_{i=1}^3n^{(i)}(v')q^i,~~~v'=\frac{Q}{M},
\end{displaymath}
and $n^{(i)}$ are three four vectors defined by
\begin{equation}
n^{(i)}(v)=\left\{v^i, \delta^{ij}+\frac{1}{1+v^0}v^iv^j\right\}.
\end{equation}
Using relations (8)-(9) we can transform the matrix element (3) to the form
containing the product of two traces over the spinor indices of both particles.
For this aim the following relations are useful:
\begin{equation}
S_{\alpha\beta}(\Lambda)u^\lambda_\beta({\bf p})=\sum_{\sigma=-s}^su^\sigma_\alpha(\Lambda{\bf p})D^s_{\sigma\lambda}(R^W_{\Lambda p}),
\end{equation}
\begin{displaymath}
\bar u_\beta^\lambda({\bf p})S^{-1}_{\beta\alpha}(\Lambda)=\sum_{\sigma=-s}^sD^{+,S}_{\lambda\sigma}
(R^W_{\Lambda p})\bar u^\sigma_\alpha(\Lambda {\bf p}).
\end{displaymath}
Substituting expressions (8), (9), (13) in Eq.(3) and using relations (17)
we obtain:
\begin{equation}
<\Psi_{n,Q}|J_\mu^{(0)}|\Psi_{n,P}>=\int\frac{d{\bf p}d{\bf q}}{(2\pi)^3}
\bar\Psi_0({\bf q})\bar u_1(0)\frac{(\hat q_1+m_1)}{\sqrt{2\epsilon_1(q)(
\epsilon_1(q)+m_1)}}S^{-1}(L_Q)\gamma_\mu S(L_P)\times
\end{equation}
\begin{displaymath}
\times\frac{(\hat p_1+m_1)}{\sqrt{2\epsilon_1(p)+m_1)}}u_1(0)\bar u_2(0)
\frac{(\hat q_2+m_2)}{\sqrt{2\epsilon_2(q)+m_2)}}S^{-1}(L_Q)\times
\end{displaymath}
\begin{displaymath}
\times S(L_P)\frac{(p_2+m_2)}{2\epsilon_2(p)(\epsilon_2(p)+m_2)}u_2(0)
\delta({\bf p_2}-{\bf q_2})\Psi_0({\bf p}),
\end{displaymath}
where
\begin{displaymath}
p_1=(\epsilon_1(p),{\bf p}), p_2=(\epsilon_2(p), -{\bf p}),
q_1=(\epsilon_1(q), {\bf q}), q_2=(\epsilon_2(q),-{\bf q}).
\end{displaymath}
Introducing the one-particle projectors on the states with definite spins
$s_{1,2}$ in the rest frame
\begin{equation}
\hat\pi_{1,2}=[u_{1,2}(0)\bar u_{1,2}(0)]=\frac{(1+\gamma^0)}{2}\frac{(1+\gamma^5\hat s_{1,2})}
{2},
\end{equation}
we can represent basic matrix element (18) in the form:
\begin{equation}
<\Psi_{n,Q}|J_\mu^{(0)}(0)|\Psi_{n,P}>=\int\frac{d{\bf p}d{\bf q}}{(2\pi)^3}
\bar\Psi_0({\bf q})\Psi_0({\bf p})\delta({\bf p_2}-{\bf q_2})\times
\end{equation}
\begin{displaymath}
\times Tr\left[\frac{(\hat q_1+m_1)}{\sqrt{2\epsilon_1(q)(\epsilon_1(q)+m_1)}}
S^{-1}(L_Q)\gamma_\mu S(L_P)\frac{(\hat p_1+m_1)}{\sqrt{2\epsilon_1(p)(\epsilon_1(p)+m_1)}}
\frac{(1+\gamma^0)(1+\gamma^5\hat s_1)}{4}\right]\times
\end{displaymath}
\begin{displaymath}
\times Tr\left[\frac{(\hat q_2+m_2)}{\sqrt{2\epsilon_2(q)(\epsilon_2(q)+m_2)}}
S^{-1}(L_Q)S(L_P)\frac{(\hat p_2+m_2)}{\sqrt{2\epsilon_2(p)(\epsilon_2(p)+m_2)}}
\frac{(1+\gamma^0)(1+\gamma^5\hat s_2)}{4}\right].
\end{displaymath}
In the next orders of the perturbative series the matrix element of electromagnetic
current keeps the general structure (20) but the form of the vertex function
(5) will be changed. The expression (20) is useful tool for the calculation
of the different order corrections to the lepton gyromagnetic factors by means
of the computer systems of analytical calculations. The system Form
\cite{Form} is employed in the calculation of the correction $\Delta g_{str}$.

\section{Nuclear structure corrections of order $(Z\alpha)^4$}

In the framework of the quasipotential method formulated in the section 2
there are two sources of the nuclear structure contributions of order
$(Z\alpha)^4$ to the gyromagnetic factors of the bound particles.
The first contribution is represented on the diagrams Fig.1 (b). The interaction
vertex of the second particle (nucleus) with electromagnetic field (large bold
circle) is determined by two form factors $F_{1,2}$.

\begin{figure}
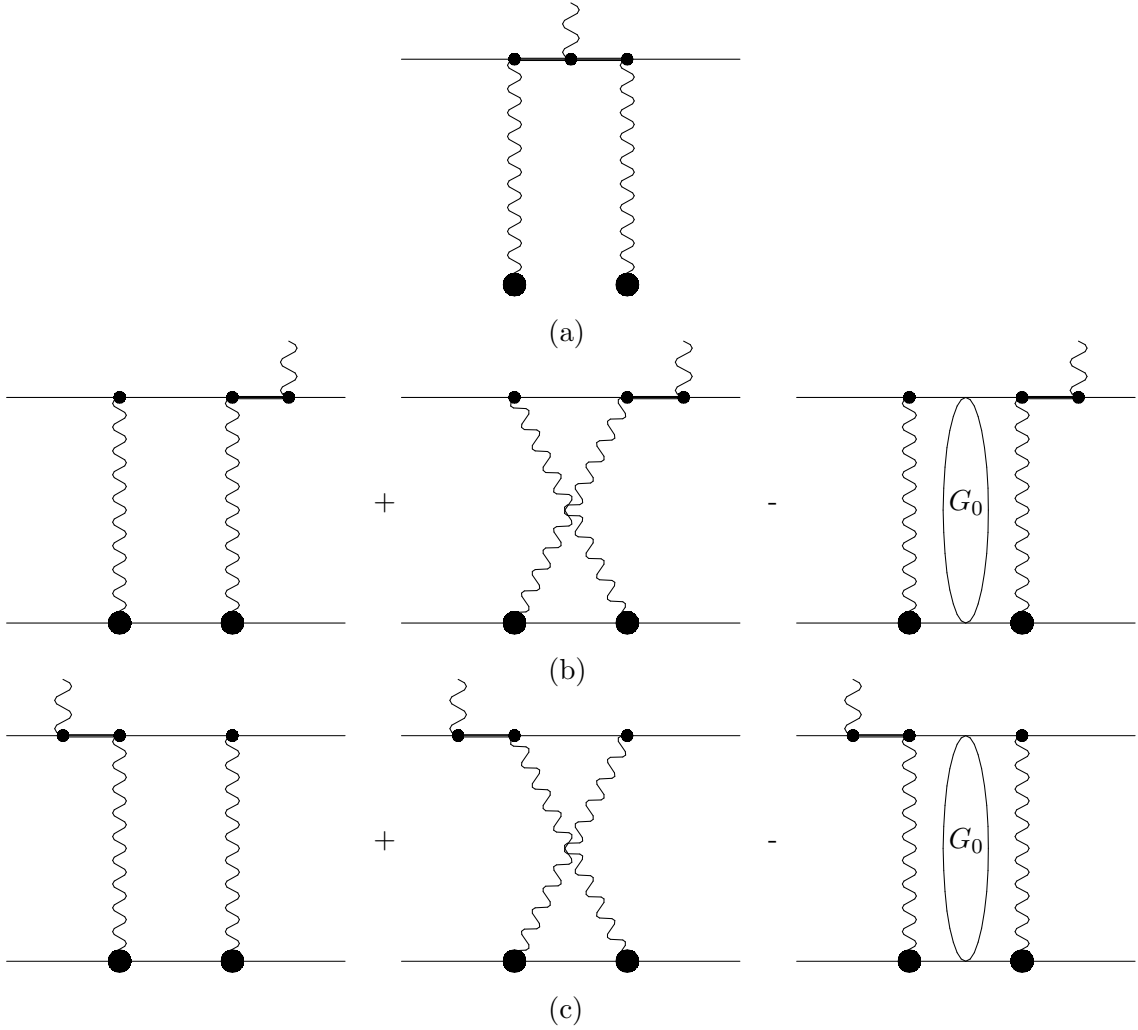

\magnitude=2000
\GRAPH(hsize=15){
\mov(0,0){\lin(3.,0)}%
\mov(0,2.){\lin(3.,0)}%
\mov(3.5,0){\lin(3,0)}%
\mov(3.5,2.){\lin(3,0)}%
\mov(7,0){\lin(3,0)}%
\mov(7,2){\lin(3,0)}%
\mov(1.,0.){\Circle**(0.2)}%
\mov(2.,0){\Circle**(0.2)}%
\mov(4.5,0.){\Circle**(0.2)}%
\mov(5.5,0.){\Circle**(0.2)}%
\mov(8,0){\Circle**(0.2)}%
\mov(9,0){\Circle**(0.2)}%
\mov(9,2){\Circle**(0.1)}%
\mov(8,2){\Circle**(0.1)}%
\mov(1.,0){\wavelin(0,2)}%
\mov(2.,0){\wavelin(0,2)}%
\mov(1,2){\Circle**(0.1)}%
\mov(2,2){\Circle**(0.1)}%
\mov(0.5,2){\Circle**(0.1)}%
\mov(0.5,2){\wavelin(0,0.5)}%
\mov(0.5,1.99){\lin(0.5,0)}%
\mov(0.5,2.01){\lin(0.5,0)}%
\mov(4.5,2){\Circle**(0.1)}%
\mov(5.5,2){\Circle**(0.1)}%
\mov(4.5,0){\wavelin(1.,2.)}%
\mov(5.5,0){\wavelin(-1.,2.)}%
\mov(4.,2){\wavelin(0.,0.5)}%
\mov(7.5,2.){\wavelin(0.,0.5)}%
\mov(4.,1.99){\lin(0.5,0)}%
\mov(4.,2.01){\lin(0.5,0)}%
\mov(7.5,1.99){\lin(0.5,0)}%
\mov(7.5,2.01){\lin(0.5,0)}%
\mov(7.5,2.){\Circle**(0.1)}%
\mov(4,2){\Circle**(0.1)}%
\mov(8.,0.){\wavelin(0,2)}%
\mov(9,0){\wavelin(0,2)}%
\mov(8.5,1.){\Ellipse(0.4)[1,5]}%
\mov(8.35,1.){$G_0$}%
\mov(8.35,4.){$G_0$}%
\mov(4.8,-0.5){(c)}%
\mov(0.,3){\lin(3.,0.)}%
\mov(0.,5){\lin(3.,0)}%
\mov(3.5,3){\lin(3,0)}%
\mov(3.5,5){\lin(3,0)}%
\mov(7,3){\lin(3,0)}%
\mov(7,5){\lin(3,0)}%
\mov(1,3){\Circle**(0.2)}%
\mov(2,3){\Circle**(0.2)}%
\mov(1,5){\Circle**(0.1)}%
\mov(2,5){\Circle**(0.1)}%
\mov(4.5,3){\Circle**(0.2)}%
\mov(5.5,3){\Circle**(0.2)}%
\mov(4.5,5){\Circle**(0.1)}%
\mov(5.5,5){\Circle**(0.1)}%
\mov(6.,5){\Circle**(0.1)}%
\mov(2.5,5){\Circle**(0.1)}%
\mov(2.5,5){\wavelin(0,0.5)}%
\mov(6,5){\wavelin(0,0.5)}%
\mov(1,3){\wavelin(0,2)}%
\mov(2,3){\wavelin(0,2)}%
\mov(4.5,3){\wavelin(1,2)}%
\mov(5.5,3){\wavelin(-1,2)}%
\mov(8,3){\wavelin(0,2)}%
\mov(9,3){\wavelin(0,2)}%
\mov(8.5,4.){\Ellipse(0.4)[1,5]}%
\mov(8,3){\Circle**(0.2)}%
\mov(9,3){\Circle**(0.2)}%
\mov(8,5){\Circle**(0.1)}%
\mov(9,5){\Circle**(0.1)}%
\mov(9.5,5){\Circle**(0.1)}%
\mov(9.5,5){\wavelin(0,0.5)}%
\mov(2,4.99){\lin(0.5,0)}%
\mov(2,5.01){\lin(0.5,0)}%
\mov(5.5,4.99){\lin(0.5,0)}%
\mov(5.5,5.01){\lin(0.5,0)}%
\mov(9,4.99){\lin(0.5,0)}%
\mov(9,5.01){\lin(0.5,0)}%
\mov(3.25,1){+}%
\mov(6.75,1){-}%
\mov(3.25,4){+}%
\mov(6.75,4){-}%
%\mov(3.5,6){\lin(3,0)}%
\mov(3.5,8){\lin(3,0)}%
\mov(4.5,6){\Circle**(0.2)}%
\mov(5.5,6){\Circle**(0.2)}%
\mov(4.5,8){\Circle**(0.1)}%
\mov(5.5,8){\Circle**(0.1)}%
\mov(5,8){\Circle**(0.1)}%
\mov(5,8){\wavelin(0,0.5)}%
\mov(4.5,6){\wavelin(0,2)}%
\mov(5.5,6){\wavelin(0,2)}%
\mov(4.5,7.99){\lin(1,0)}%
\mov(4.5,8.01){\lin(1,0)}%
\mov(4.8,2.5){(b)}%
\mov(4.8,5.5){(a)}%
}
\caption{Generalized two-particle vertex function $\Gamma_\mu$:
the nuclear structure corrections of order $(Z\alpha)^5$.
Bold line denotes the negative energy part of the lepton propagator.}
\end{figure}

Let us consider the contributions of the diagrams Fig.1 (b). The trace over
the lepton spinor indices for the first diagram Fig.1 (b) entering in Eq. (20)
takes the form:
\begin{equation}
T_{11}=\frac{1}{64m_1^2}Tr\Biggl\{(1+\gamma^0)(1+\gamma^5\hat s_1)(\hat q_1+m_1)
\gamma^0\left[\hat v'+1-(1-\gamma^0)\right]\gamma^\nu\times
\end{equation}
\begin{displaymath}
\times\Lambda^-({\bf p'_1})\gamma^0\gamma^\mu
\left[\hat v+1-(1-\gamma^0)\right]\gamma^0(\hat p_1+m_1)\Biggr\},
\end{displaymath}
\begin{equation}
\Lambda^-({\bf p'_1})=\frac{1}{2m_1}\gamma^0\left[m_1(\gamma^0-1)+\hat p+
\hat\Delta\left(1-\frac{m_1}{2M}\right)\right], {\bf p}'_1={\bf p}_1+{\bf\Delta},
~p=(0,{\bf p}).
\end{equation}
Similar contribution of the second diagram Fig.1 (b) is equal
\begin{equation}
T'_{11}=\frac{1}{64m_1^2}Tr\Biggl\{(1+\gamma^0)(1+\gamma^5\hat s_1)(\hat q_1+m_1)
\gamma^0\left[\hat v'+1-(1-\gamma^0)\right]\gamma^\mu\times
\end{equation}
\begin{displaymath}
\times\Lambda^-({\bf q'_1})\gamma^0\gamma^\nu
\left[\hat v+1-(1-\gamma^0)\right]\gamma^0(\hat p_1+m_1)\Biggr\},
\end{displaymath}
\begin{equation}
\Lambda^-({\bf q'_1})=\frac{1}{2m_1}\gamma^0\left[m_1(\gamma^0-1)+\hat q-
\hat\Delta\left(1-\frac{m_1}{2M}\right)\right], {\bf q}'_1={\bf q}_1-{\bf\Delta},~
q=(0,{\bf q}).
\end{equation}
The trace over the spinor indices of the second particle (nucleus) which is
equal for both diagrams Fig. 1 (b) has the following structure:
\begin{equation}
T_{21}=\frac{1}{64m_2^2}Tr\Biggl\{(1+\gamma^0)(1+\gamma^5\hat s_2)(\hat q'_2+m_2)
\gamma^0\left[\hat v'+1-(1-\gamma^0)\right]\gamma^0\times
\end{equation}
\begin{displaymath}
\times\left[\gamma^\nu F_1+\frac{i}{2m_2}F_2\sigma^{\nu\lambda}k_\lambda\right]
\left[\hat v+1-(1-\gamma^0)\right](\hat p'_2+m_2)\Biggr\}.
\end{displaymath}
After the contraction of the expressions (21) and (23) with (25) over the Lorentz
index $\nu$ ($\mu\to i$ is the vector index) we obtain that the sum of the trace
products can be written in the form:
\begin{equation}
[\mathstrut\bm{\sigma}_1\times{\bf\Delta}]F_1\left(1-\frac{m_1}{m_1+m_2}\right)+
[\mathstrut\bm{\sigma}_2\times{\bf\Delta}]\left(F_1+2F_2\right)
\frac{m_1}{m_1+m_2}.
\end{equation}
Substituting sequentially Eq.(26) in Eq.(20) and then in Eqs.(3) and (2) and
calculating the $rot_{{\bf\Delta}}$ we keep only the contributions
to the lepton gyromagnetic factor (the particle 1). The nuclear structure
correction takes the form (${\bf s_1}\to<\mathstrut\bm{\sigma}_1>$):
\begin{equation}
\Delta g^{str}_1=\frac{4}{3}m_1^2R^2_N(Z\alpha)^4\left(1-
4\frac{m_1}{m_2}\right),
\end{equation}
where the following expansion of the Dirac form factor $F_1$ at small ${\bf k}^2$ is
used:
\begin{equation}
F_1({\bf k}^2)=1-\frac{{\bf k}^2}{6}R^2_N.
\end{equation}
The expression (27) contains not only the correction of order $(Z\alpha)^4$
but also the recoil effects of the first order over the ratio $m_1/m_2$. The same order
correction as (27) appears in the second order of the perturbation theory.
The contribution of the second order of the perturbative series to the matrix
element (3) has the form:
\begin{equation}
2<\Psi_n|{\bf J}(0)|\delta\Psi_n>=2<\Psi_n|{\bf J}\tilde G\Delta V|\Psi_n>=
2\sum_{m}\frac{<\Psi_n|{\bf J}(0)|\Psi_m><\Psi_m|\Delta V|\Psi_n>}{E_n-E_m},
\end{equation}
where $\Psi_n$ are the Coulomb wave functions of Eq.(10). The reduced Coulomb
Green function (RCGF) $\tilde G$ doesn't contain the free two-particle
Green function because its contribution cancels the term which appears from
the half of last iteration contribution $\Gamma^{(2)}$ in Eq.(7). The nuclear structure effects
in Eq.(29) are determined by the terms of the quasipotential
$\Delta V\sim R^2_N$ as it follows from Eq. (28). To calculate the matrix
element in Eq.(29) we need to know the form of relativistic correction of
order ${\bf p}^2/m_1^2$ including recoil effects in the vertex operator
$\Gamma_\mu$ which was found in Refs.\cite{F1,F2}. Accounting the form of the
$1S$ state RCGF in the coordinate representation \cite{IK}
\begin{equation}
\tilde G_{1S}(r,0)=\frac{Z\alpha\mu^2}{4\pi}\frac{2e^{-x/2}}{x}\left[2x(\ln x+
C)+x^2-5x\right],
\end{equation}
($C$=0.5772156649... is the Euler constant) we can express necessary nuclear structure
correction in the g-factor of the bound lepton in the second order of the
perturbative series as follows:
\begin{equation}
\Delta g_2^{str}=-\frac{8}{3}R^2_N
\pi Z\alpha\int\Psi_{1S}({\bf r})\frac{{\bf p}^2}{3m_1^2}\left(1+
\frac{m_1}{2m_2}\right)d{\bf r}\tilde G_{1S}(r,0)|\Psi_n(0)|=
\end{equation}
\begin{displaymath}
=\frac{4}{3}R^2_Nm_1^2(Z\alpha)^4\left(1-\frac{7m_1}{2m_2}\right).
\end{displaymath}
Summing two corrections (27) and (31) we obtain total contribution of the
nuclear structure effects of order $(Z\alpha)^4$ with the account recoil
correction:
\begin{equation}
\Delta g^{str}_{l,(Z\alpha)^4}=
\frac{8}{3}R^2_Nm_1^2(Z\alpha)^4\left(1-\frac{15m_1}{4m_2}\right).
\end{equation}
Numerical values (32) for several hydrogenic ions are presented in
Table \ref{t2}.

\begin{table}
\caption{\label{t2} Nuclear structure corrections in the
g-factor of the bound electron and muon of orders $(Z\alpha)^4$ and $(Z\alpha)^5$.}
\bigskip
\begin{ruledtabular}
\begin{tabular}{|c|c|c|c|c|c|c|}   \hline
$Z$ &  $R_N (fm)$ & $b_{NS}[\times 10^{-9}]$  & \multicolumn{2}{|c|}
{electron $\Delta g^{str}$} &\multicolumn{2}{|c|}{muon $\Delta g^{str}$}\\
\cline{4-5}\cline{6-7}
 &  &\cite{K}   &  $(Z\alpha)^4[\times 10^{-9}]$ & $(Z\alpha)^5[\times 10^{-12}]$ &
 $(Z\alpha)^4[\times 10^{-4}]$ & $(Z\alpha)^5[\times 10^{-6}]$  \\  \hline
7  & 2.54 & 0.79 & 0.79 & 0.13 & 0.33 & -0.17 \\  \hline
8  & 2.737 & 1.56 & 1.56 & 0.26& 0.65 & -0.42 \\  \hline
9  & 2.90  & 2.80 & 2.80 & 0.53& 1.17 & -0.89 \\ \hline
10 & 2.992 & 4.54 & 4.54 & 0.95 &1.90 & -1.66 \\ \hline
12 & 3.08 & 9.98 & 9.97 & 2.49  & 4.19 &-4.54 \\ \hline
15 & 3.191 & 26.14 & 26.13 & 8.10& 11.03 & -15.50 \\  \hline
18 & 3.423 & 62.37 & 62.37 & 22.80 & 26.39 & -47.88 \\  \hline
24 & 3.643 & 223.29 & 223.28 &117.13  & 94.68 &-244.35  \\  \hline
32 & 4.088 & 888.63 & 888.60 &550.02  &377.57 & -1461.07\\  \hline
\end{tabular}
\end{ruledtabular}
\end{table}

\section{Spin dependent nuclear structure corrections of order
$(Z\alpha)^4$}

In general case,
the matrix element of the electromagnetic current for the particle of arbitrary spin $S$
is determined by means of $(2S+1)$ form factors (charge, magnetic,
quadruple, et al.). When the magnetic moment of simple atomic
systems is studied it may be possible to take into account the form factors of the
minimal multipolarity describing the distributions of the electric charge and
magnetic moment. The one-particle matrix element $J_\mu$ of the electromagnetic current
operator between states with momenta $p$ and $q$ can be written as follows:
\begin{equation}
J_{\mu}=\bar U({\bf p})\left\{\Gamma_\mu F^D_1+\frac{1}{2m}\Sigma_{\mu\nu}
k^\nu F^P_2\right\}U({\bf q}).
\end{equation}
The wave function $U({\bf p})$ of a particle with arbitrary spin entering
in Eq.(33) can be presented in the form (see, for instance \cite{KMS,KP}):
\begin{equation}
U=\left(\xi\atop\eta\right)=\left(\xi^{\alpha_1\alpha_2...\alpha_p}_{\dot\beta_1
\dot\beta_2...\dot\beta_q}\atop\eta^{\beta_1\beta_2...\beta_q}_{\dot\alpha_1\dot
\alpha_2...\dot\alpha_p}\right),~~~p+q=2S,
\end{equation}
where the spin-tensors $\xi$, $\eta$ are symmetrical in upper and lower indices.
For the particle of half integer spin p=S+1/2, q=S-1/2. In the case of integer
spin p=q=S. The Lorentz transformation of the spinors $\xi$ and $\eta$
can be written in the form \cite{KP,RF}:
\begin{equation}
\xi=\exp\left(\frac{{\bf\Sigma}\mathstrut\bm{\delta}}{2}\right)\xi_0,~~~\eta=\exp\left(
-\frac{{\bf\Sigma}\mathstrut\bm{\delta}}{2}\right)\xi_0,
\end{equation}
where the direction of the vector $\mathstrut\bm{\delta}$ coincides with the velocity of
the particle, $th\delta=v$. The generator of the Lorentz transformation
${\bf\Sigma}$ is equal:
\begin{equation}
{\bf\Sigma}=\sum_{i=1}^p\mathstrut\bm{\sigma}_i-\sum_{i=p+1}^{p+q}\mathstrut\bm{\sigma}_i,
\end{equation}
and $\mathstrut\bm{\sigma}_i$ acts on the ith index of the spinor $\xi_0$ as follows:
\begin{equation}
\mathstrut\bm{\sigma}_i\xi_0=(\mathstrut\bm{\sigma}_i)_{\alpha_i\beta_i}(\xi_0)_{...\beta_i...}.
\end{equation}
In the standard representation, which is introduced in analogy with the spin
1/2, the free particle wave function (34) can be written with the accuracy
$(v/c)^2$ in the form:
\begin{equation}
U({\bf p})=\left(\left[1+\frac{({\bf\Sigma}{\bf p})^2}{8m^2}\right]\xi_0
\atop \frac{{\bf\Sigma}{\bf p}}{2m}\xi_0\right).
\end{equation}
The components of the matrix $\Sigma_{\mu\nu}$ in (33) are the generators of
the boosts and rotations \cite{KP,RF}:
\begin{equation}
\Sigma_{n0}=\left(\begin{array}{cc}
\Sigma_n & 0\\
0 & -\Sigma_n\end{array}\right),~~~\Sigma_{mn}=-2i\epsilon_{mnk}
\left(\begin{array}{cc}
s_k  &  0\\
0  &  s_k\end{array}\right),~~{\bf s}=\frac{1}{2}\sum_{i=1}^{2S}\mathstrut\bm{\sigma}_i.
\end{equation}
Considering the contribution of the diagrams Fig.1 (b) to the matrix element
(20) let us point out that with desired accuracy the bispinor contraction
over line of the second particle (nucleus)
\begin{equation}
\bar U_2({\bf q}_2){\cal B}_2U_2({\bf p}_2)F_1(k^2)\approx\left(1-\frac{1}{6}k^2R_N^2
\right)
\end{equation}
(where the matrix ${\cal B}_2$ is the generalization for the matrix $\beta_2$
used in the case of the spin 1/2 particles) doesn't contain any other
additional factors connected with the nuclear spin. So, the form of the
correction (27) remains unchanged with the account the first order recoil
effect. But the correction of the second order over the lepton-to-proton mass
ratio contains the spin dependent terms. Indeed, using Eqs.(33), (38)-(39) we
have the following bispinor contraction for both particles (see the diagrams
Fig.1 (b) where the nucleus interacts with external electromagnetic field):
\begin{displaymath}
\frac{e_2}{2m_2}\bar u_1({\bf q}_1)\alpha^i_1u_1({\bf p}_1)F_1(k^2)
\bar U_2({\bf q}_2){\cal B}_2\left[{\cal A}_2^i\frac{1}{2}\left
(1-{\cal B}_2\right){\cal\bf A}_2+{\cal\bf A}_2\frac{1}{2}\left(1-{\cal B}_2
\right){\cal A}_2^i\right]U({\bf p}_2)\approx
\end{displaymath}
\begin{equation}
\approx -\frac{Ze}{2m_2}\frac{m_1}{m_1+m_2}i\left[\mathstrut\bm{\sigma}_1
\times{\bf\Delta}\right]K_{S_2}\frac{1}{6}{\bf k}^2R_N^2,
\end{equation}
where
\begin{equation}
{\cal \bf A}=\left(\begin{array}{cc}
0 & {\bf\Sigma}\\
{\bf\Sigma} & 0\end{array}\right),~~
{\cal B}=\left(\begin{array}{cc}
I& 0\\
0& -I\end{array}\right),
\end{equation}
\begin{equation}
K_{s_2}=\frac{<{\bf\Sigma}_2^2>}{3}=\Biggl\{{\frac{4s_2}{3},
~s_2~{\rm is~integer~nuclear~spin}\atop\frac{4s_2+1}{3},~s_2~{\rm
is~half~integer~nuclear~spin}}.
\end{equation}
The following commutation relations are useful:
\begin{equation}
\left[\Sigma_i,\Sigma_j\right]=4i\epsilon_{ijk}s_k,~~\left[\Sigma_i,s_j\right]=
i\epsilon_{ijk}\Sigma_k.
\end{equation}
Other part of the contribution of order
$(Z\alpha)^4m_1^2/m_2^2$ is determined by relativistic effects in the vertex operator
$\Gamma^{(0)}$ which were obtained in Ref.\cite{MF1} on the basis of relations
(34)-(39) in the form:
\begin{equation}
\Delta\Gamma^{(0)}_{rel}=\frac{e_1}{2m_1}\mathstrut\bm{\sigma}_1\left\{-\frac{{\bf p}^2}
{3m_1^2}\left(1+\frac{m_1}{2m_2}\right)+\frac{{\bf p}^2}{m_2^2}\left[1-K_{S_2}
\left(1+\frac{Z}{3}\right)-\frac{Z-1}{6}\right]\right\}.
\end{equation}
Calculating the contribution of the second order of the perturbative series by
means of Eqs.(29) and (44) and taking into account the relation (41) we obtain
the following total value of the nuclear structure correction:
\begin{equation}
\Delta g^{str}_{l,total}=
\frac{8}{3}m_1^2R_N^2(Z\alpha)^4\left\{1-\frac{15m_1}{4m_2}+
\frac{m_1^2}{4m_2^2}\left[6K_{S_2}+Z+29\right]\right\}.
\end{equation}
It is helpful to remark that two terms in Eqs.(41) and (45) proportional
to $K_{s_2}\cdot Z$ will be mutually cancelled in the correction (46).
The expression (46) is the generalization of Eq.(32) for the case of the nuclear
spin dependent contribution which contains also the squared ratio of the particle
masses $m_1/m_2$. Numerical values of the third addendum in Eq.(46) for different
$Z$ and $s_2$ are compared with the correction of order
$(Z\alpha)^6m_1^2R_N^2$ \cite{GS} and represented in Table \ref{t3}.

\begin{table}
\caption{\label{t3} Spin dependent nuclear structure corrections in the g-factor
of the bound electron and muon of order $(Z\alpha)^4$: $\Delta g^{str}_l=
\frac{2}{3}R^2_Nm_1^2(Z\alpha)^4
\frac{m_1^2}{m_2^2}\left[6K_{S_2}+Z+29\right]$.}
\bigskip
\begin{ruledtabular}
\begin{tabular}{|c|c|c|c|c|}   \hline
$Z$ &  $R_N (fm)$ & $s_2$ & electron [$\times 10^{-16}$] & muon [$\times 10^{-6}$]  \\  \hline
7  & 2.54 & 1/2 & 0.12 & 0.02  \\
   &      & 1  &  0.13 & 0.02  \\   \hline
8  & 2.737 & 0 & 0.17 & 0.03  \\
   &       & 5/2& 0.27 & 0.05  \\   \hline
11  & 2.94  & 3/2 & 0.53 & 0.10 \\
    &       & 3  & 0.63  & 0.12  \\
    &       & 4 & 0.71 &  0.13  \\   \hline
13 & 3.035 & 5/2 & 0.94 & 0.17 \\
   &       & 3   & 0.97  & 0.18 \\  \hline
24 & 3.643 & 7/2 & 5.28 &0.97   \\  \hline
\end{tabular}
\end{ruledtabular}
\end{table}

\section{Nuclear structure corrections of order $(Z\alpha)^5$}

The nuclear structure corrections of order $(Z\alpha)^5$ in the g-factor of the
bound lepton are specified by the vertex function $\Gamma^{(2)}$ in Eq.(7) and
shown in Fig.2. There is need to take into account that the corrections
of such order contain also additional small factor $m_1/\Lambda$
($\Lambda$ is the form factor parameter) which lead to the decrease this
contribution as compared with Eq.(32). Nevertheless just as for the g-factor
of the bound electron in hydrogenic ion with sufficiently high $Z$, so for
gyromagnetic factor of the bound muon the value of this correction can
increase essentially and become important for modern experiments.

The two-photon exchange diagrams relevant to the $\Gamma^{(2)}$ can be divided
into two parts. In the Feynman amplitudes of the first part the interaction
of the lepton with external field occures between two exchange interactions.
We consider these two-photon exchanges in the limit of static nucleus
neglecting recoil corrections when the nucleus interacts with the photons
via the Dirac form factor $F_1$. The interaction of the lepton with two
exchanged Coulomb photons is plotted in Fig. 2 (a).
The bispinor contraction over the lepton line with consideration
for the interaction terms (7) incorporates negative energy parts of the lepton
propagators shown in Fig.2 (a) by bold line. The trace over the spinor indices
of the first particle entering in Eq. (20) can be written as follows:
\begin{equation}
T_{12}=\frac{1}{16}Tr\Bigl\{(1+\gamma^0)(1+\gamma^5\hat s_1)
(\hat q_1+m_1)[\hat v'+1-(1-\gamma^0)]\gamma^0\times
\end{equation}
\begin{displaymath}
\times\left[\epsilon_1({\bf k}+{\bf\Delta})-k^0+\gamma^0(\hat k+\hat\Delta-m_1)\right]
\gamma^0\gamma^\lambda\left[\epsilon_1({\bf k})-k^0+\gamma^0(\hat k-m_1)\right]
[\hat v+1-(1-\gamma^0)](\hat p_1+m_1)\Bigr\}
\end{displaymath}
\begin{displaymath}
\approx 4m_1^2(\epsilon_1(k)-m_1)[\mathstrut\bm{\sigma}_1\times{\bf\Delta}].
\end{displaymath}
Then we can write the contribution of the diagram Fig.2 (a) to the g-factor of
the bound lepton in the form ($\epsilon_1(k)=\sqrt{k^2+m_1^2}$):
\begin{equation}
\Delta g^{(1)}_{l,(Z\alpha)^5}=-\frac{29m_1\mu^3(Z\alpha)^5}
{32\pi}\int_0^\infty\frac{dk}{k^2}\frac{(F_1^2-1)(\epsilon_1(k)-m_1)}
{(\epsilon_1(k)+m_1)^2\epsilon_1(k)^2}.
\end{equation}
The scale of the integration momenta in Eq.(48) is of order of the lepton
mass. In spite of the infrared finiteness of the appeared expression in the
leading order over $m_1/m_2$
we make the subtraction of the point-like nuclear contribution in
Eq.(48). To obtain correct energy spectrum
in the case of point-like nucleus it is necessary
to consider the bound state effects and keep the relative motion momenta
of the particles ${\bf p}$, ${\bf q}$ (see the discussion, for example,
in Ref.\cite{EGS}). Another part of the two-photon exchange diagrams
which lead to the nuclear structure
corrections of order $(Z\alpha)^5$ is presented in Fig.2 (b), (c).
The trace calculations over the spinor indices and the contraction over the
Lorentz indices can be performed for these diagrams in the same manner as
for previous contributions. To illustrate the general structure of intermediate
expressions in this case we write the trace product regarding to the first
diagram in Fig.2 (b):
\begin{equation}
Tr\Bigl\{(1+\gamma^0)(1+\gamma^5\hat s_1)(\hat q_1+m_1)(\hat v'+1)
\gamma^\lambda\Lambda^-(-{\bf\Delta})\gamma^0\gamma^\mu(\hat p_1+\hat k+m_1)
\gamma^\nu(\hat v+1)(\hat p_1+m_1)\Bigr\}\times
\end{equation}
\begin{displaymath}
\times Tr\Bigl\{(1+\gamma^0)(1+\gamma^5\hat s_2)(\hat q_2+m_2)(\hat v'+1)
\gamma^\mu(m_2\gamma^0-\frac{1}{2}\hat\Delta-\hat k+m_2)
\gamma^\nu(\hat v+1)(\hat p_2+m_2)\Bigr\}F_1(k)F_1(k+\Delta).
\end{displaymath}
Total contribution of all diagrams in Fig. 2 (b),
(c) in the trace product of Eq.(20) is the following:
\begin{equation}
m_2F_1^2\Biggl\{-\frac{50}{3}k_0^4[\mathstrut\bm{\sigma}_1\times{\bf\Delta}]-32m_1^2k_0^2
[\mathstrut\bm{\sigma}_1\times{\bf k}]-\frac{32}{3}k^4[\mathstrut\bm{\sigma}_1\times{\bf\Delta}]+
\end{equation}
\begin{displaymath}
+8k^4[\mathstrut\bm{\sigma}_1\times{\bf k}]-\frac{14}{3}k^2k_0^2
[\mathstrut\bm{\sigma}_1\times{\bf\Delta}]+16k^2k_0^2[\mathstrut\bm{\sigma}_1\times{\bf k}]\Biggr\}.
\end{displaymath}
But contrary to earlier considered expressions for the matrix elements
of electromagnetic current (2) which are linear with respect to
$[\mathstrut\bm{\sigma}_1\times{\bf\Delta}]$ these diagrams involve also the integral
of the form $\int d^4k[\mathstrut\bm{\sigma}_1\times{\bf k}] F(k,\Delta)$. Its
calculation includes the shift of the integration variable
${\bf k}\to{\bf k}-\frac{\bf\Delta}{2}$ and the calculation of the
$rot_{\bf\Delta}$ from the integration function. After rotating the $k^0$
contour of the Feynman loop integration we integrate over four dimensional
Euclidean space using the relation
\begin{equation}
\int d^4k=4\pi\int k^3dk\int_0^\pi\sin^2\phi d\phi,~~k^0=k\cos\phi.
\end{equation}
The integration over the angle variable $\phi$ can be done analytically.
After that we can present the sum of the contributions
shown in Fig.2 (b), (c) to the lepton g-factor in the form:
\begin{equation}
\Delta g^{(2)}_{l,(Z\alpha)^5}=-\frac{4}{3}
\frac{m_2\mu^3(Z\alpha)^5}{\pi}\int_0^\infty\frac{dk}{k}\times
\end{equation}
\begin{displaymath}
\times\left\{(F_1^2-1)\left[25I_3+16I_1+16k^2I_4+31I_2+8k^2\frac{d}{dm_1^2}
(I_1-I_2)\right]-16F'_1(0)\frac{\mu}{m_1m_2k}\right\},
\end{displaymath}
\begin{equation}
I_1=\frac{1}{\pi}\int_{-1}^1\frac{\sqrt{1-x^2}dx}{(k^2+4m_1^2x^2)(k^2+4m_2^2x^2)}=
\frac{\omega_2-\omega_1}{4k^3(m_2^2-m_1^2)}, \omega_i=\sqrt{k^2+4m_i^2},
\end{equation}
\begin{equation}
I_2=\frac{1}{\pi}\int_{-1}^1\frac{x^2\sqrt{1-x^2}dx}{(k^2+4m_1^2x^2)(k^2+4m_2^2x^2)}=
\frac{k(m_1^2-m_2^2)+m_2^2\omega_1-m_1^2\omega_2}{16k(m_2^2-m_1^2)m_1^2m_2^2},
\end{equation}
\begin{equation}
I_3=\frac{1}{\pi}\int_{-1}^1\frac{x^4\sqrt{1-x^2}dx}{(k^2+4m_1^2x^2)(k^2+4m_2^2x^2)}=
\frac{2m_1^2m_2^2(m_1^2-m_2^2)+k^2(m_1^4-m_2^4)+k(m_2^4\omega_1-m_1^4\omega_2)}
{64m_1^4m_2^4(m_1^2-m_2^2)},
\end{equation}
\begin{equation}
I_4=\frac{1}{\pi}\int_{-1}^1\frac{(1-x^2)^{3/2}dx}{(k^2+4m_1^2x^2)^2(k^2+4m_2^2x^2)}=
\end{equation}
\begin{displaymath}
=\frac{1}{16k^5\omega_1(m_1^2-m_2^2)^2}\left[-k^4+k^2(6m_2^2-2m_1^2+\omega_1
\omega_2)+4(2m_1^4-6m_1^2m_2^2+m_2^2\omega_1\omega_2)\right].
\end{displaymath}
The contribution of the iteration term in the vertex operator $\Gamma^{(2)}$
(7) was taken into account in Eq.(52). We make also necessary subtraction
of the point-like nuclear contribution in Eq.(52) because in the low momentum
region there exists logarithmic infrared divergence.
To obtain numerical results on the
basis of the expressions (48) and (52) the dipole parameterization for
the Dirac form factor $F_1$ is used:
\begin{equation}
F_1(k^2)=\frac{1}{\left(1+\frac{k^2}{\Lambda^2}\right)^2},~~\Lambda=\frac{\sqrt{12}}{R_N}.
\end{equation}
The total contributions of order $(Z\alpha)^5$ to the $\Delta g_{str}$
for the electron and muon presented in Table \ref{t2} have different signs
because the corrections (48) and (52) have opposite signs and their numerical
values differ considerably.

\section{Conclusion}

The calculation of the nuclear structure corrections (the nuclear size
corrections) of orders $(Z\alpha)^4$
and $(Z\alpha)^5$ in the electron and muon g-factors is performed in this
work on the basis of the quasipotential method. Numerical results of
obtained corrections (32), (46), (48), (52) are presented in Tables \ref{t2}
and \ref{t3}. We
kept the same values of the nuclear charges and nuclear charge radii as
in Ref.\cite{PSSL}. It was demonstrated that considered corrections can
be successfully derived in the framework of the quasipotential method used earlier
in Refs.\cite{F1,F2} for the study the radiative and recoil effects of orders
$O(\alpha^2)$, $O(\alpha^3)$. The matrix element of electromagnetic current
operator is obtained in new form (20) which is convenient for the calculation
of different quantum electrodynamical corrections of high order over $\alpha$
and $m_1/m_2$ by means of such systems of analytical calculations
as the Form, Reduce etc. Obtained results
for the corrections of order $(Z\alpha)^4$ coincide with the calculations
in Refs. \cite{K,GS}. New recoil contributions $(Z\alpha)^4m_1/m_2$,
$(Z\alpha)^4(m_1/m_2)^2$ and the correction of order $(Z\alpha)^5$
in the g-factor of the bound lepton are
numerically less important to the present than the leading order contribution.
Their role
can be significantly increased with the growth of experimental accuracy for the
hydrogenic ions with sufficiently high $Z$. All obtained analytical relations
and numerical results for the correction $\Delta g^{str}$ in the $1S$ state can be
extended to other $S$ states of simple atoms. It should be pointed out that
for muonic hydrogenic atoms the role of the nuclear structure corrections
of orders $(Z\alpha)^4$ and $(Z\alpha)^5$ including recoil effects in the muon
g-factor increases by some orders of the magnitude as compared with the g-factor
of the bound electron.
The theoretical expression for the gyromagnetic factor of the bound
muon in the hydrogen-like atom accounting relativistic, radiative, recoil and
structure corrections of the fourth order has the form \cite{KIS}:
\begin{equation}
g_\mu(H)=2\Bigl\{1-\frac{(Z\alpha)^2}{3}\left[1-\frac{3}{2}\frac{m_\mu}
{m_N}\right]-\frac{(1+Z)(Z\alpha)^2}{2}\left(\frac{m_\mu}{m_N}\right)^2+
\frac{\alpha(Z\alpha)^2}{2\pi}\left[\frac{1}{6}-\frac{1}{3}\frac{m_\mu}{m_N}
\right]-
\end{equation}
\begin{displaymath}
-\frac{(Z\alpha)^4}{12}-\frac{2\alpha(Z\alpha)^2}{\pi}\left[\frac{2}{9}
\ln\left(\frac{2Z\alpha m_\mu}{m_e}\right)-\frac{5}{27}+\frac{m_e^2}
{2(Z\alpha m_\mu)^2}+O(m_\mu/m_N)\right]+\frac{4}{3}m_\mu^2R_N^2(Z\alpha)^4
\Bigr\}.
\end{displaymath}
In the case of the muonic hydrogen numerical value of the
nuclear structure correction of order $(Z\alpha)^4$ is equal $1.11
\times 10^{-9}$ whereas the total value of the corrections to the gyromagnetic
factor $g_\mu(H)$ following from Eq.(58) is $(1-15490\times 10^{-9})$.
So, an enlargement of experimental region to the g-factor of
the bound muon in the muonic hydrogenic atoms which is studied in this work,
and an enhancement of experimental accuracy can lead to more precise
determination of several fundamental physical constants such as the
proton charge radius and the muon-to-ptoton mass ratio.

\acknowledgments

The author is grateful to D.Ebert, R.N.Faustov and V.O.Galkin
for useful discussions. The work was performed under the financial
support of the Russian Fond for Basic Research (grant No. 04-02-16085).


\begin{thebibliography}{99}
\bibitem{MT}P.J.Mohr, B.N.Taylor, Rev. Mod. Phys. {\bf 72} (2000) 351.
\bibitem{TB}T.Beier, S.Djekic, H.Haffner et al. Nucl. Instr. and Methods
in Phys. Research {\bf B205} (2003) 15.
\bibitem{BT}T.Beier, Phys. Rep. {\bf 339} (2000) 79.
\bibitem{F1}R.N.Faustov, Phys. Lett. {\bf B33} (1970) 422.
\bibitem{F2}R.N.Faustov, Nuovo Cimento. {\bf 69} (1970) 37.
\bibitem{G}H.Grotch, Phys. Rev. Lett. {\bf 24} (1970) 39.
\bibitem{GH}H.Grotch, R.A.Hegstrom, Phys. Rev. {\bf A4} (1971) 59.
\bibitem{CO}F.E.Close, H.Osborn, Phys. Lett. {\bf 34} (1971) 400.
\bibitem{EG}M.I.Eides, H.Grotch, Ann. Phys. {\bf 260} (1997) 191.
\bibitem{BCS}S.A.Blundell, K.T.Cheng, J.Sapirstein, Phys. Rev. {\bf A5}
(1997) 1857.
\bibitem{PSSL}H.Persson, S.Salomonson, P.Sunnergren, I.Lindgren,
Phys. Rev. {\bf A56} (1997) R2499.
\bibitem{VS}V.M.Shabaev, Can. J. Phys. {\bf 76} (1998) 907.
\bibitem{K}S.G.Karshenboim, Phys. Lett. {\bf A266} (2000) 380.
\bibitem{Y}V.A.Yerokhin, P.Indelicato, V.M.Shabaev, Can. J. Phys. {\bf 80}
(2002) 1249.
\bibitem{KIS}S.G.Karshenboim, V.G.Ivanov, V.M.Shabaev, JETP {\bf 93} (2001) 478.
\bibitem{GS}D.A.Glazov, V.M.Shabaev, Phys. Lett. {\bf A297} (2002) 408.
\bibitem{MF1}A.P.Martynenko, R.N.Faustov, JETP {\bf 93} (2001) 471.
\bibitem{TR}J.S.Tiedeman, H.G.Robinson, Phys. Rev. Lett. {\bf 39} (1977) 602.
\bibitem{WPK}F.G.Walther, W.D.Phillips, D.Kleppner, Phys. Rev. Lett. {\bf 28}
(1972) 1159.
\bibitem{LR}D.J.Larson, N.F.Ramsey, Phys. Rev. {\bf A9} (1974) 1543.
\bibitem{JR}C.E.Johnson, H.G.Robinson, Phys. Rev. Lett. {\bf 45} (1980) 250.
\bibitem{H}H.Haffner, T.Beier, N.Hermanspahn et al., Phys. Rev. Lett.
{\bf 85} (2000) 5308.
\bibitem{Werth}J.L.Verdu, S.Djekic, T.Valenzuela et al., Can. J. Phys. {\bf 80}
(2002) 1233.
\bibitem{F3}R.N.Faustov, Ann. Phys. {\bf 78} (1973) 176.
\bibitem{MF2}A.P.Martynenko, R.N.Faustov, Theor. Math. Phys. {\bf 64} (1985)
765.
\bibitem{Form}J.A.M.Vermaseren, E-preprint math-ph/0010025, 2000.
\bibitem{IK}V.G.Ivanov, S.G.Karshenboim, JETP {\bf 109} (1996) 1219.
\bibitem{KMS}I.B.Khriplovich, A.I.Milstein, R.A.Sen'kov, Phys. Lett. {\bf A221}
(1996) 370.
\bibitem{KP}I.B.Khriplovich, A.A.Pomeransky, JETP {\bf 113} (1998) 1537.
\bibitem{RF}Yu.B.Rumer, A.I.Fet, Group theory and Quantum Fields, M., Nauka,
1977.
\bibitem{EGS}M.I.Eides, H.Grotch, V.A.Shelyuto, Phys. Rep. {\bf 342}
(2001) 63.
\end{thebibliography}
\end{document}